\documentclass[final,5p,times]{elsarticle}
\usepackage{graphicx}
\graphicspath{{/home/dean/Documents/LaTeX/latex-documents/visio_graphics/}}
\usepackage{amsmath}
\usepackage{amssymb}
\usepackage{bbm}
\usepackage[utf8]{inputenc}
\usepackage[T1]{fontenc}

\usepackage{caption}

\usepackage{ifthen}
\usepackage{enumitem}

\newboolean{APPENDIX}
\setboolean{APPENDIX}{true} 

\newcommand{\quotes}[1]{``#1''}

\newcommand{\antiinstanton}{(anti\nobreakdash-)instanton}

\newcommand{\imagi}{\mathsf{i}}

\newcommand{\su}{\mathsf{SU}}

\usepackage{color}

\renewcommand{\vec}[1]{\mathbf{#1}}

\newcommand{\Lqcd}{Lattice QCD}
\newcommand{\lqcd}{lattice QCD}

\journal{SCCG 2017}

\begin{document}
\sloppy
\begin{frontmatter}


\title{QCDVis: a tool for the visualisation of Quantum Chromodynamics (QCD) Data}

\author[label1,label2]{Dean P.Thomas}
\author[label3]{Rita Borgo}
\author[label1]{Robert S.Laramee}
\author[label2]{Simon J.Hands}
\address[label1]{Department of Computer Science, College of Science, Swansea University, United Kingdom}
\address[label2]{Department of Physics, College of Science, Swansea University, United Kingdom}
\address[label3]{Informatics Department, Kings College London, London, United Kingdom}

%\author[label1,label2]{author 1}
%\author[label3]{author 2}
%\author[label1]{author 3}
%\author[label2]{author 4}
%
%\address[label1]{Department 1}
%\address[label2]{Department 2}
%\address[label3]{Department 3}

\begin{abstract}
	Quantum chromodynamics, most commonly referred to as QCD, is a relativistic quantum field theory for the strong interaction between subatomic particles called quarks and gluons. The most systematic way of calculating the strong interactions of QCD is a computational approach known as lattice gauge theory or lattice QCD. Space-time is discretised so that field variables are formulated on the sites and links of a four dimensional hypercubic lattice. This technique enables the gluon field to be represented using $3 \times 3$ complex matrices in four space-time dimensions. Importance sampling techniques can then be exploited to calculate physics observables as functions of the fields, averaged over a statistically-generated and suitably weighted ensemble of field configurations. In this paper we present a framework developed to visually assist scientists in the analysis of multidimensional properties and emerging phenomena within QCD ensemble simulations. Core to the framework is the use of topology-driven visualisation techniques which enable the user to segment the data into unique objects, calculate properties of individual objects present on the lattice, and validate features detected using statistical measures.  The framework enables holistic analysis to validate existing hypothesis against novel visual cues with the intent of supporting and steering scientists in the analysis and decision making process.  Use of the framework has lead to new studies into the effect that variation of thermodynamic control parameters has on the topological structure of lattice fields.
\end{abstract}


\end{frontmatter}

%\linenumbers


\section{Introduction}
\label{sec::introduction}

\noindent
Gravity and electromagnetism are two fundamental forces which we directly observe and experience. 
Physicists however identify four fundamental forces in nature: electromagnetic force, gravitational force, weak interaction, and strong interaction.  The weak and strong interaction forces work at the atomic level: the strong interaction holds the nucleus together and the weak interaction is responsible for radioactive decay.  Strong force along with electromagnetism and weak interaction, underpins all of nuclear physics and the vast array of phenomena from the microscopic to the cosmological level which we all experience.  With respect to analytical frameworks the gravitational force is governed by Einstein's theory of general relativity while the strong interaction, electromagnetic force, and weak interaction are described by quantum field theory~\textemdash{} theoretical frameworks for constructing quantum mechanical models of subatomic particles in particle physics.
%Quantum chromodynamics, most commonly referred to as QCD, is
%a relativistic quantum field theory used to describe strong interaction, e.g. the strong nuclear force and one of the four fundamental forces in nature: 

Quantum chromodynamics (QCD) is a relativistic quantum field theory used to describe the strong interaction force.  QCD acts at nuclear level where it predicts the existence of force-carrier particles called gluons and quarks.  Gluons and quarks are confined in the protons and neutrons that make up everyday matter and transmit the strong force between particles of matter.

While fundamentals of QCD can be deemed simple and elegant~\cite{Wilczek:2000ih} due to its compact Lagrangian, its phenomenology is characterised by multidimensional properties and emerging phenomena with complex dynamics making analysis an elaborate task. 
QCD simulations are performed on a discrete space-time lattice as an approximation of the theory known as lattice Quantum Chromodynamics (\lqcd{}).  Due to the complexity of the simulation data, it is crucial to extract possible features and provide semantics to become information.  This work is therefore driven by a desire to support scientists in the validation, inspection, and understanding of the properties of \lqcd{} data.
To achieve this goal we contribute a system to support physicists in the process of:
\begin{itemize}[noitemsep]
	\item validation and analysis of simulated QCD data sets;
	\item generation of visual designs that provide novel insight into the structure and evolution of phenomena represented by simulated QCD data;
	\item simultaneous comparison of both quantitative and qualitative aspects of QCD phenomena.  
\end{itemize}

In the development of our framework we leverage the strength of topology-driven analysis, successfully introduced in different research domains and range of applications all sharing a common goal~\textemdash{} to achieve an efficient and compact description of the basic shapes co-existing in an arbitrary set of data.

\section{Background}

\noindent
In this section we introduce the relevant background topics required to understand the techniques used in the QCDVis framework.  We give a brief overview of \lqcd{}, covering topics relevant to this application paper.  \Lqcd{} is a widely researched and expansive field of theoretical physics, hence, a more comprehensive review of the topic is included in the Appendix.  We limit the application background to visualisation in \lqcd{}, for a more general discussion of visualisation techniques in the physical sciences we refer the reader to the survey paper by Lipşa et al.~\cite{CGF:CGF3184}.

\subsection{Domain background}
\label{sec::domain_background}

\noindent
Quantum Chromodynamics defines a theory of interaction between subnuclear quark and gluon particles.  Matter that is made up of quarks bound together by the exchange of massless gluon particles is referred to Hadronic matter.  The Hadron group can be subdivided into two categories; tri-quark systems are baryons and quark-anti-quark pairs are mesons.  Protons and neutrons, collectively known as nucleons, are members of the baryon class of particle.

Quarks and gluons always exist in \quotes{colour neutral} states, making it near to impossible to observe them in isolation~\cite{creutz1983quarks}.  This phenomena, known as \emph{colour-confinement}, sees the amount of energy required to separate bound particles increase exponentially with distance.  For this reason QCD is often modelled using a discrete computational model of quark-gluon systems known as \lqcd{}.

The lattice structure, first proposed by Kenneth Wilson~\cite{WIL74}, is a hyper-torus in four space-time dimensions.  \Lqcd{} uses Euclidean representation of space-time where space and time are treated as equivalent.  This allows the structure to be thought of as being composed of hypercubic cells.  Periodic, or \emph{translationally-invariant}, boundary conditions in all four dimensions mean that all vertices on the lattice are treated equally, with no cell being sited on a boundary.

Quarks exist at vertices, or \emph{sites}, on the lattice with integer indices.  Each site has four \emph{link variables} represented by $\su(n)$ matrices modelling the gluon potential between neighbouring sites in the $x, y, z$ and $t$ directions.  The value of $n$ is used to represent the number of charge colours used in the gauge theory, where true QCD uses $n = 3$.  The work described here uses a simplified theory where $n = 2$ to allow certain thermodynamic control parameters to varied.  \emph{Chemical potential} ($\mu$) is one such parameter that affects the relative balance of positive and negative charge in the system, which is thought to be characterised by topological variations in the scalar fields.

Various fields exist on the lattice, defined as closed loops from a site.  Mathematically these are represented as a path ordered matrix multiplication of link variables.  The smallest closed loop in a 2D plane is referred to as a \emph{plaquette}.  In this work we are primarily interested in the plaquette fields, Polyakov loop fields, and topological charge density field.  Plaquette fields are the averages of three plaquettes, categorised as space-like ($xy$, $xz$, $yz$) or time-like ($xt$, $yt$, $zt$) plaquettes.  The Polyakov loop is a three dimensional field computed as a product of all link variables in the time direction in path order, looping at the boundary.  Topological charge density is a loop in all four space-time directions that predicts the existence of pseudo-particles named instantons~\cite{belavin1975pseudoparticle}.

Instantons are one of the primary observables that physicists working in a \lqcd{} study.  Their name comes from the fact that they have a finite existence in the time-dimension and are able to appear and disappear, unlike real particles.  Instantons are generally able to persist in the data even when subject to a lattice based noise reduction technique called cooling~\cite{Teper:1985rb}.  However, overuse of the algorithm can destroy the instanton-anti-instanton pairs that are intended to be measured \cite{Kenny2010}.  Due to the discrete nature of the lattice and the fact the size of an instanton can vary, it is possible to shrink objects until they disappear.  The effect of annihilation of an observable by overcooling is referred to an object \quotes{falling through the lattice} on account of the discrete spacing between lattice sites.

\subsection{Visualisation in Lattice QCD}

\noindent
The traditional use of visualisation in \lqcd{} is often limited to dimension independent forms such as line graphs to plot properties of simulations as part of a larger ensemble.  Visualisation of a space-time nature is less frequently observed but does occasionally occur in literature on the subject.  Surface plots were used by Hands~\cite{hands1990lattice} to view 2D slices of data and give a basic understanding of energy distributions in a field. 

Application of visualisation of QCD in higher dimensions first appears in the form of three dimensional plots~\cite{Feurstein1997} of instantons showing their correlation to other lattice observables.  Primarily, these were monopole loops, instantons that persist throughout the entire time domain of a simulation.  Prior to visualisation, simulations are subjected to iterative cooling to remove noise whilst keeping the desired observables intact.  Instantons are visualised using their plaquette, hypercube, and L\"uscher definitions, each of which make it possible to view instantons and monopole loops.  Visualisation is a useful tool in this case to confirm that monopole loops and instantons are able to coexist in a simulation, a fact that can help theoretical physicists understanding of the QCD vacuum structure.

A more detailed visualisation, making use of established computer graphics techniques, was performed by Leinweber~\cite{leinweber2000visualizations} as a way of conveying to the physics community that visualisation of large data sets would help their understanding of QCD.  An area of \lqcd{} investigated is the view that instanton-anti-instanton pairs can attract and annihilate one another during cooling phases. Topological charge density, also referred to as topological charge density, and Wilson action, also referred to as \emph{action} measurements are used to show separate spherical instantons and anti-instantons in the lattice hyper-volume.  Using animation it is possible to see that instantons in vicinity of anti-instantons deform and merge.  Leinweber explains the need for sensible thresholds when discarding isosurfaces that have action densities below a critical value as their disappearance can be incorrectly interpreted as them falling through the lattice.

More recently DiPierro et al.~\cite{di2007visualization, MAS001} produced several visual designs as part of a larger project with the aim of unifying various active \lqcd{} projects.  Expectations of the work were that visualisations could be used to identify errors with the simulation process, along with increasing understanding of \lqcd{}.  Many existing file structures used within the \lqcd{} community could be processed and output using the VTK, this could then be rendered using a number of pre-existing visualisation tools including VisIt~\cite{childs2013visit} and Paraview~\cite{Ahrens2005}.  In addition, the evolution of instantons over Euclidean time spans could be rendered into movies.  A web front end application and scripting interface was also provided that allowed users to set up an online processing chain to analyse simulations on a large scale.  Despite the apparent success of the work it would appear that it is no longer maintained and many of the links present with the project literature are inaccessible.

%-------------------------------------------------------------------------
\subsection{Topology Driven Visualisation}

\noindent
Topological descriptions of data aim to achieve an efficient depiction of the basic components and their topological changes as a function of a physical quantity significant for the specific type of data.   Contour trees, a direct derivative of Morse Theory, provide a compact data structure to encapsulate and represent the evolution of the isocontours of scalar data, or height function, as the height varies and relate these topological changes to the criticalities of the function. 
	
De Berg et al.~\cite{DeBerg1997a} made the observation that an isosurface is the level set of a continuous function in 3D space suggesting that a whole contour could be traced starting from a single element~\textemdash{} a seed.  An entire contour can be computed for the desired isovalue from the seed that is bounded by two super-nodes at the critical points of the contour.  Carr et al.~\cite{Carr2003} investigated the use of contour trees in higher dimensional datasets, whilst also improving upon the algorithm proposed by Tarasov and Vyalyi~\cite{tarasov1998construction}.  They introduced the concept of augmented contour trees, an extension that added non-branching vertices to provide values for isosurface seeding. Separate isosurfaces can be generated for each non-critical value identified in the contour tree, separately colour coded and manipulated by the end user.  This enables users to identify regions of interest and distinguish them accordingly~\cite{carr2004topological}.  
		
Chiang et al.~\cite{chiang2005simple} introduce a modified contour tree construction algorithm that improves processing time by only sorting \emph{critical vertices} in the merge tree construction stage.  This vastly increases processing speeds in very large data sets.  However, as observed by the authors, critical vertices are difficult to identify in data with dimensionality of four or more.  Further increases in speed are offered by storing multiple seeds for the monotone path construction algorithm~\cite{takahashi1995algorithms} used to generate surfaces.  This results in an increased storage overhead but with the advantage that a surface does not require complete re-extraction each time the isovalue is changed.
		
Limitations in the merge tree computation phase of the contour tree algorithm prevent it from correctly handling data that is periodic in nature.  However, the Reeb graph~\cite{cole2003loops}, a direct descendent of the contour tree, can be used to compute the topology of scalar data with non-trivial boundaries.  As with the contour tree, the Reeb graph can be applied to models of any dimension provided it is represented on a simplicial mesh~\cite{pascucci2007robust}.  Tierny et al.\ proposed an optimized method for computing Reeb graphs using existing contour tree algorithms by applying a procedure that they called loop surgery~\cite{tierny2009loop}.  This method is able to take advantage of the speed of contour trees with a minimal overhead in complexity.  The Reeb graph has also been used to assist in the design of transfer functions in volume visualisation~\cite{fujishiro2000volume}, by assigning opacity based upon how many objects were obscured by nested surfaces and their proximity to the edge of a scalar field.

In comparison to previous tools in \lqcd{} that rely upon static visualisation, using existing scientific software packages, our system is a bespoke dynamic environment designed collaboratively with physicists.  The inclusion of topological structures in the system allows us to present unique interpretations of the data which to our knowledge has not been carried out in the domain before.  As the system is dynamic, the user is able to explore data in real-time by adjusting visualisation parameters such as isovalue or to interact with the data at level captured by the scalar topology.  

Unique features of QCDVis directly influenced a study by Thomas et al.~\cite{Thomas2017}, where topological features of the Polyakov loop are examined as chemical potential and temperature are varied.  Initial visual analysis of the polyakov loop data showed a complex interaction between isosurfaces with scalar values across the field range.  When the surfaces were viewed with the additional context of the Reeb graph capturing the topology, it was questioned if there could potentially be a bias in the number and structure of the objects in the field.  In order to capture details of the topology as quantitative data, aspects of the Reeb graph and geometric properties were computed for multiple data sets.
		
\section{Research Aims}

\noindent
Scientists are particularly interested in examining behaviours of observables computed via path integral equations, for field configurations generated using the unphysical value $N_c=2$ and with the addition of a term proportional to a parameter $\mu$ known as the \emph{chemical potential}~\cite{Hands:2006ve, Hands:2010gd, cotter2013towards}. When $\mu>0$, the Monte Carlo sampling is biased towards configurations with an excess of quarks $q$ over anti-quarks $\bar q$.  As a result exploration of systems with quark density $n_q>0$, relevant for understanding the ultra-dense matter found within neutron stars, is made possible.  Study of ``Two Colour QCD'' is an exploratory step in this direction: in the presence of quarks, the action $S$ is supplemented by an additional contribution $S_q(U;\mu)$ from quarks which is in general non-local and extremely expensive to calculate.

For the physical case $N_c=3$, it can be shown that $S_q(\mu)=S_q^*(-\mu)$, where the star denotes complex conjugate, so that for $\mu\not=0$ the action is neither real nor positive definite, making importance sampling inoperable. For $N_c=2$, $S_q(\mu)$ is real and the problem disappears.

For any relativistic system with $\mu\not=0$, the $t$ direction is distinguished, because $n_q>0$ implies particle world-lines are preferentially oriented in one time-like direction (a world-line is the trajectory of a particle through space-time, and is oriented for a particle such as a quark carrying a conserved charge).  An interesting question is how this implied space-time anisotropy is reflected in the properties of the topological fluctuations.  Previous studies~\cite{han2011a} suggest instanton sizes shrink as $\mu$ increases but it is also possible that their shape and relative displacement from neighbouring (anti-)instantons may have different characteristic distributions in $t$ and $x,y,z$ directions.

\begin{figure*}[!htb]
	%\centering
	\includegraphics[width=\textwidth]{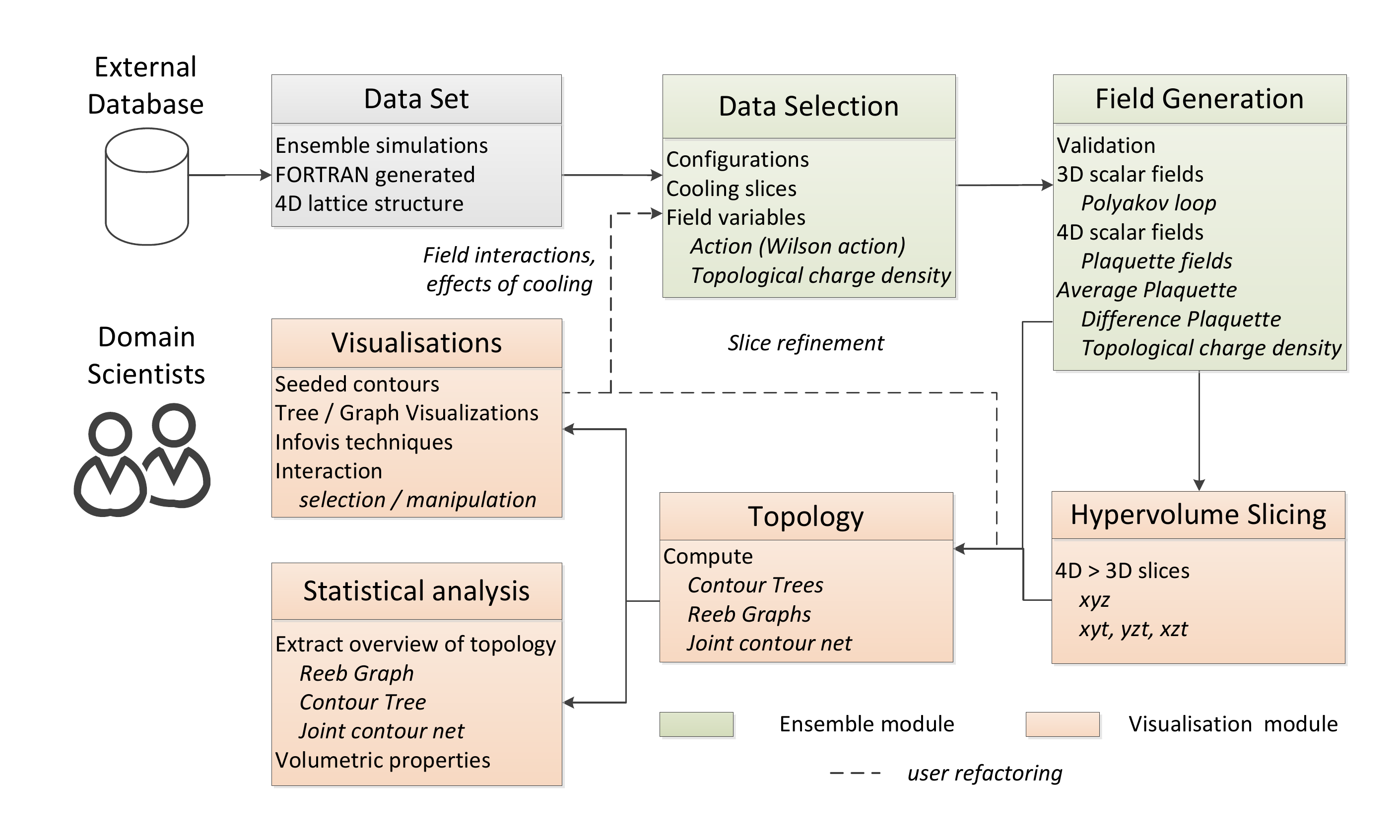}
	\caption{QCDVis Visual Analytics pipeline.}
	\label{fig::QCDVisToolWorkFlow}
\end{figure*}

Existing attempts at visualisation of QCD data have so far concentrated on single aspects of the analysis~\cite{leinweber2000visualizations}, or have relied upon static visualisations~\cite{di2012visualization}.  At present no framework exists capable of supporting the scientist through the entire analytical pipeline from beginning to end and back through the use of interactive visualisations.  Our framework has been developed to fill this gap.  QCDVis (Figs.~\ref{fig::QCDVisToolWorkFlow},~\ref{fig::ensembleTool},~\ref{fig::QCDVis_main}) has been designed with the user in mind. The framework addresses the fundamental aspects of data validation, visualisation and estimation of QCD objects and observables, manipulation and traversal of data.  Core to the analysis is support for identification of discrete objects and their characteristics using topological structures data such as the Reeb graph.  This enables research questions to be tackled in a quantitative fashion. The following sections provide detailed descriptions of each aspect of our framework and its implementation.

\section{QCDVis}

\noindent
QCDVis has been designed as an interactive environment to enable physicists to visually explore their data and its underlying topology.  Figure~\ref{fig::QCDVisToolWorkFlow} summarizes the key stages of the visual analytic pipeline upon which our system is constructed.  Along with 3D rendered views of user selected scalar fields, we also include data representations that are commonly found in the QCD domain.  Histograms, in particular, are firmly established in physics community for identifying underlying trends in data.  Therefore, they are included to allow physicists to gain an overview of their data and promote a focus on the parts of the data they find most interesting.

In order to explain the functioning of the system we first define some terminology specific to \lqcd{}:

\begin{description}[noitemsep]
	\item[Action] A function of the lattice variables computed by completing closed loops on the lattice.  In this work we use the term to indicate the \emph{Wilson action}, defined by unit loops on the lattice referred to as plaquettes.
	\item[Configuration] A unique lattice representing the Quark-Gluon field at a specified step along the simulation Markov chain.
	\item[Cooling] A form of algorithm used by physicists used to remove noise and stabilise output from \lqcd{} simulations.  %There are several cooling algorithms, each with differing methods for smoothing the data.
	\item[Ensemble] A collection of configurations with common thermodynamic control parameters, for example chemical potential ($\mu$). 
	\item[Euclidean space-time] A four dimensional manifold with three real coordinates representing Euclidean space, and one imaginary coordinate, representing time, as the fourth dimension.
	\item[Plaquette] A unit closed 2D loop around links on the lattice, starting and ending at the same vertex.
	\item[Space-like plaquette] A plaquette existing in the $xy$, $xz$ or $yz$ planes.
	\item[Time-like plaquette] A plaquette existing in the $xt$, $yt$ or $zt$ planes.
\end{description}

The QCDVis framework currently consists of two main modules; the first processes ensemble data and is used to transform raw lattice configurations into discrete scalar fields representing a number of lattice field operators.  The second, working at the configuration level, is used to visualise and probe these fields using topological methods.  Each module can work independently or can be used together to semi-autonomously probe the ensemble with respect to its topological features.  In this scenario, the ensemble tool is used to invoke the visualisation and analysis application with given sets of configurations.

\subsection{Ensemble module}
\label{sec::ensemble_module}

%-------------------------------------------------------------------------
\begin{figure}[!htb]
	\centering
	\includegraphics[width=\columnwidth]{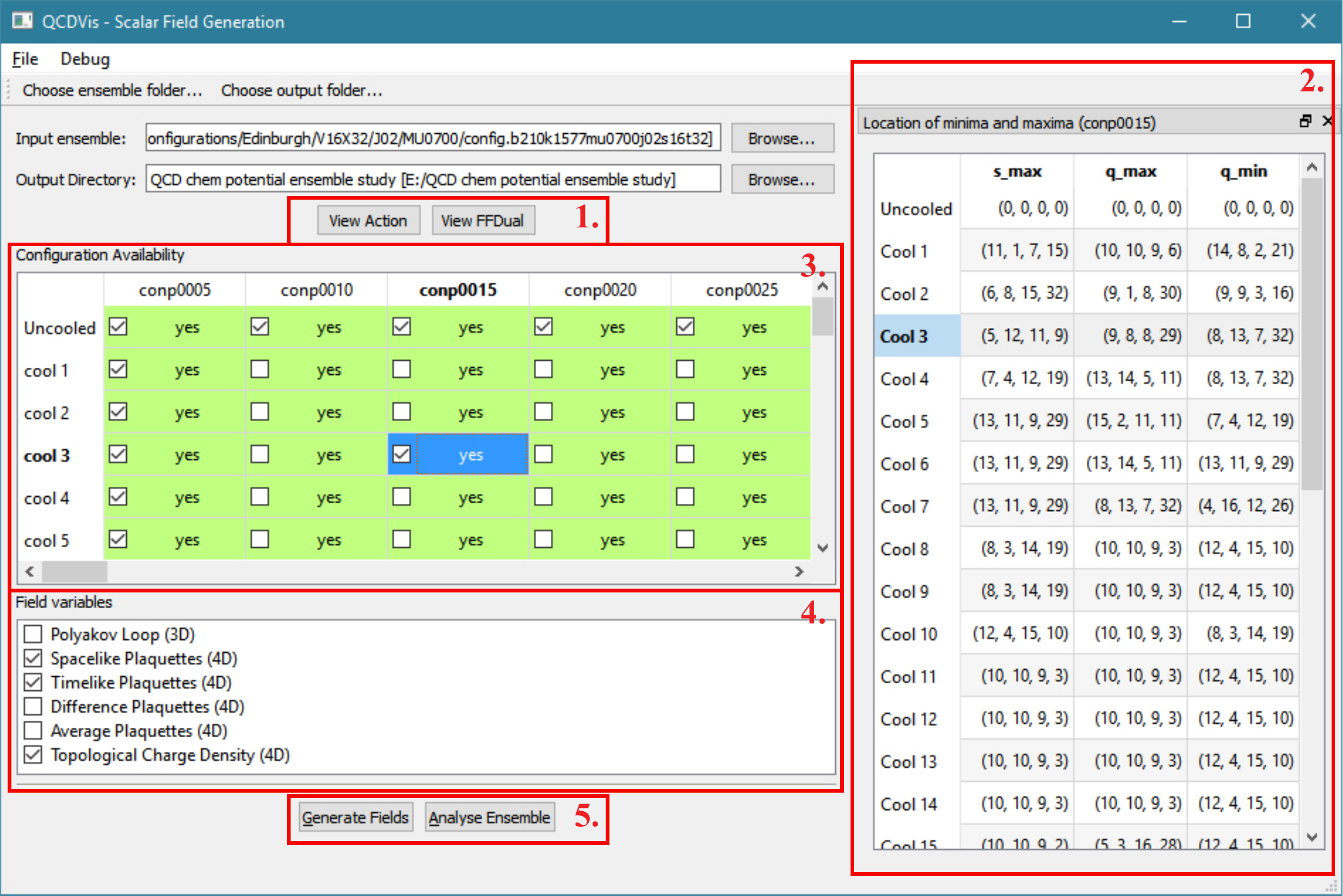}
	\caption{Ensemble interface.  Physicists can select multiple cooling slices, configurations and field variables of interest.}
	\label{fig::ensembleTool}
\end{figure}

\noindent
Figure~\ref{fig::ensembleTool} shows the interface of the ensemble computation module made up of our main components discussed below.  \Lqcd{} simulations are stored as 4D array-like structures referred to as lattice configurations.  These are not the direct observables and require processing to generate data for visualisation purposes.  The resultant data is 4D, sliced into 3D volumes along a user defined axis.
		
\paragraph{Ensemble cooling histogram generation (Fig.~\ref{fig::ensembleTool}.1)}  Upon selecting an ensemble for inspection, physicists can view underlying patterns under cooling by computing graphs of the action and topological charge density statistics (e.g., Wilson action and topological charge density respectively).  This can steer them towards interesting features in the derived scalar fields.
	
\paragraph{Location of field maxima and minima (Fig.~\ref{fig::ensembleTool}.2)}  The cooling program is able to detect the position in space-time of minima and maxima of the action and topological charge density fields as it is cooled.  This data is available for each cooling stage for each configuration and can be called up by the user from the main interface.  Physicists can use this data to determine the probable existence of (anti-)instantons in the topological charge density field, signified by corresponding minima and maxima in both fields.  In QCDVis we include this data to provide an additional method for determining interesting stages of cooling in each configuration.

\paragraph{Configuration selection (Fig.~\ref{fig::ensembleTool}.3)}  The vast quantity of configurations created for an ensemble under cooling mean that it is not viable for the physicists to view all possible configuration/cooling combinations.  Additionally, it is unusual for a domain physicist to retain configurations early in the cooling cycle, as the configurations are too noisy for use in calculations.  Therefore, physicists can select exactly which configurations they intend to further investigate using information gathered from generated histograms and localization of maxima and minima (see above).  Singular configurations can be selected by checking the relevant cell in the grid but we also offer the convenience of selecting all configurations across a level of cooling or within a configuration.
	
\paragraph{Field variable selection (Fig.~\ref{fig::ensembleTool}.4)}  At present the system is able to calculate a range of field variables including topological charge density, polyakov loops, plaquettes, and fields derived from the averages of time-like and space-like plaquettes.  Most lattice fields require the calculation of the space-like and time-like plaquettes in order to create the final scalar field.  In order to optimize the computation of these values, they are computed and cached on demand.

Differing field variables may be required depending on the nature of the study being conducted; hence, we provide a selection of variables that are calculated at runtime.  For example, use of the Polyakov loop field is a basic method of detecting \quotes{breaking of symmetry} on the lattice.  The topological charge density, average plaquette, and difference plaquette fields can all be used to predict the presence of instanton observables.

\paragraph{Field  variables calculations from configuration data (Fig.~\ref{fig::ensembleTool}.5)}  
Data is stored on a four dimensional Euclidean lattice with the $\su(2)$ matrices (denoted $\mu, \nu, \rho, \sigma$) assigned to edges in the $x$, $y$, $z$, and $t$ dimensions.  Field strength variables are defined by moving around the lattice from a given origin in closed loops.  

To calculate different lattice fields we complete closed loops from a given lattice site in the $\mu, \nu, \rho, \sigma$ directions.  Movements across the lattice in a positive direction require multiplication by the relevant $\su(2)$ link variable.  Equivalently, movement in a negative direction requires a multiplication by the matrix in its adjoint, or conjugate transpose, form.  Hence, to complete a unit loop in the $xy$ plane a path-ordered multiplication of two link variables ($\mu, \nu$) and the two link variables in their adjoint form ($\mu^\dagger, \nu^\dagger$) is performed.  
		
The main field variables of interest are topological charge, Polyakov loops, space-like and time-like plaquettes, and their average/difference (as shown in Fig.~\ref{fig::ensembleTool}.4). Topological charge density for example is a loop around all four dimensions from an origin point.  Thus values are a multiplicative combination of three spatial plaquettes and three time-like plaquettes.  The adjoint variations of each plaquette, representing a loop in reverse, are subtracted from these to form the \emph{difference plaquettes}.

\subsection{Visualisation module}
\label{sec::visualisation_module}

\begin{figure*}[!htb]
	\centering
	\includegraphics[width=\textwidth]{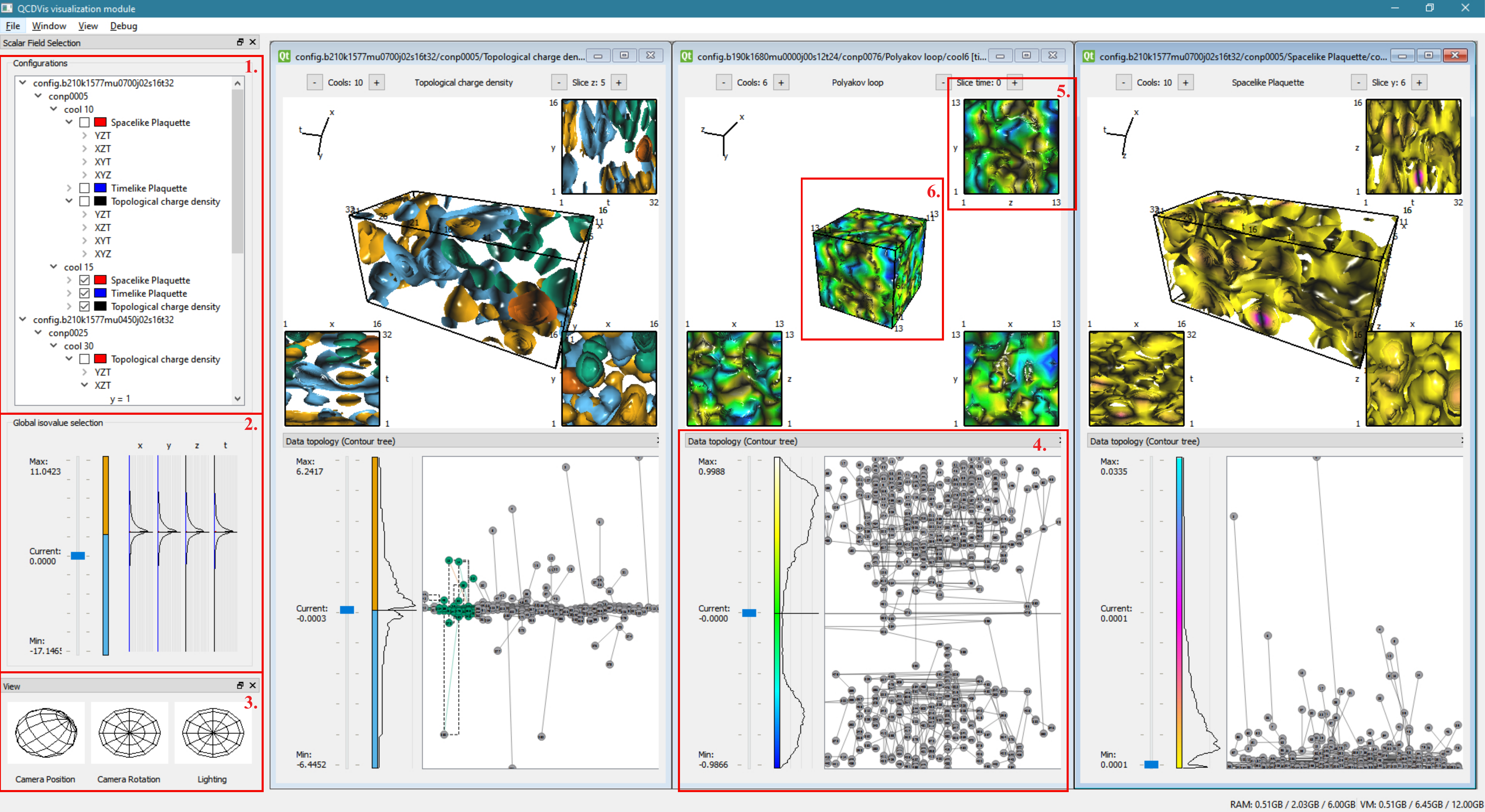}
	
	\caption{QCDVis Visualisation Interface. QCDVis enables physicists to slice and view multiple configurations and variables from different ensembles in parallel. Leftmost and rightmost visualisation showing topological charge density and space-like plaquettes of different samples from same configuration, central visualisation showing Polyakov loops from a sample extracted from a different ensemble.}
	\label{fig::QCDVis_main}
\end{figure*}

\noindent
Figure~\ref{fig::QCDVis_main} shows an overview of the visualisation module which is composed of six main sections discussed below.

\paragraph{Scalar Field Selection (Fig.~\ref{fig::QCDVis_main}.1)}  QCDVis allows multiple configurations and field variables to be viewed in parallel, many of which are 4D.  Each opened configuration is accessible through a hierarchy, the root of which is the source ensemble.  The next level of the hierarchical view is the configuration followed by the relevant cooling slice.

For fields that are a 4D in nature, all 3D sub-volumes are computed and cached when first opening the file.  An additional layer of the hierarchy sorts each of the 3D scalar fields according to the axis of slicing.  The user can view multiple scalar fields by double clicking the relevant item in the tree.  We therefore make no assumptions of how the user intends to interact with the data but, instead enable the user to choose combinations of fields based upon their own requirements.  This makes QCDVis adaptable to viewing data concurrently from a number of points-of-view:
\begin{itemize}[noitemsep]
	\item Multiple ensembles can be viewed to gain an overview of how simulation parameters, such as chemical potential, on the structure of objects.
	\item Configurations in the same ensemble can be viewed at different steps along the generating Markov chain.
	\item The effect of cooling on the lattice can be observed by viewing the same configuration at differing levels of cooling.
	\item Comparisons can be made between differing lattice variables by visual comparison of the field data.
	\item Objects can be viewed in both space-like (e.g. $xyz$) and time-like (e.g. $xzt$) orientations, this can give extra information about the structure of lattice observables in space time.
\end{itemize}

\paragraph{Global isovalue interactions (Fig.~\ref{fig::QCDVis_main}.2)}  The user is able to interact with the data globally or on an individual configuration basis.  At the global scope, we provide an overview of selected configurations using a histogram of the number of objects at each isovalue across the entire 4D data set.  Histograms are available for each of the space-time axes as it is expected that the space-like and time-like signatures will vary.  The user can also set a global isovalue using a slider control, enabling them to understand correspondences in differing configurations or lattice operators.

\paragraph{Global view interactions (Fig.~\ref{fig::QCDVis_main}.3)}  In order to visualise data from the same point-of-view we enable the user to interact globally with all open scalar fields.  For rotation we provide the user with controls using the concept of arc-ball~\cite{Shoemake1985} rotations, this gives the freedom of movement for trackball input devices. A floating camera orbits the outside of the scalar field with the radius of the sphere it sits upon modifiable using a scroll wheel.

A number of fixed cameras can also be placed by the user; these exist at each corner of the bounding box.  Additionally, a second arc-ball widget enables fine adjustment of the camera, initially fixed upon the centre of the data.  The user can also move the relative position of the lighting using a third arc-ball control.

\begin{figure*}[!htb]
	\centering
	\includegraphics[width=\textwidth]{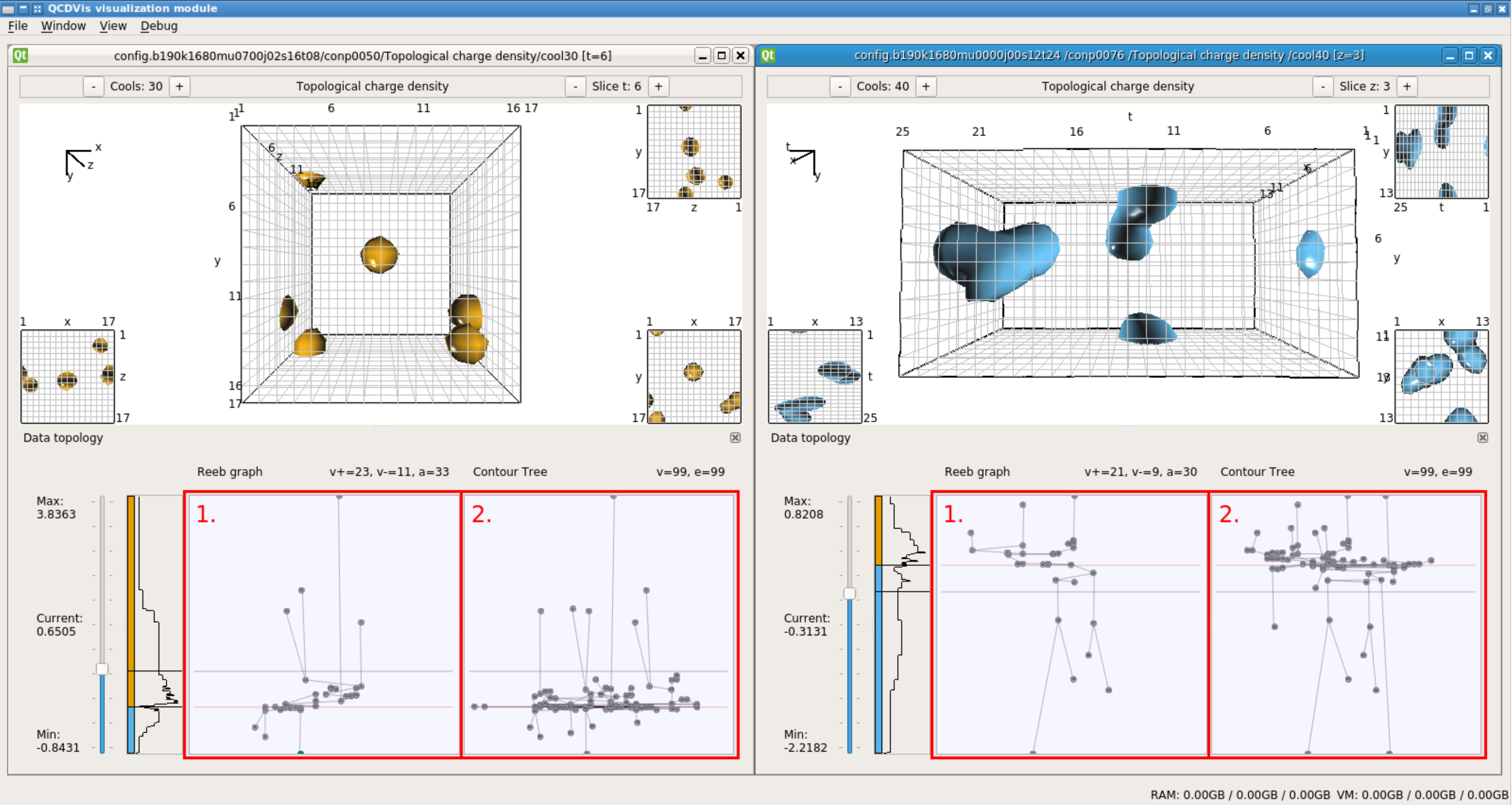}
	\caption{Left: Two objects crossing on the periodic $x$ and $y$ boundaries are joined by two non-periodic objects in a time-slice of topological charge density.  Right: An object crossing a periodic boundary on the $t$ axis, a second object crosses on the $y$ boundary.  Both visualisations are shown alongside their Reeb graph (1.) and the contour Tree (2.) representations to illustrate the difference in these topological structures when dealing with periodic boundaries.}
	\label{fig:periodicBoundaries}
\end{figure*}

\paragraph{Data topology overview and isovalue selection (Fig.~\ref{fig::QCDVis_main}.4)}  Scalar fields can be explored using a localised isovalue range.  The user can steer towards regions of  interest in the domain via either a contour tree or Reeb graph (Fig.~\ref{fig:periodicBoundaries}) representation of the domain. 
Selection in the contour tree or Reeb graph (Fig.~\ref{fig::QCDVis_main}.4) are reflected in the flexible-isosurface view, as shown in the leftmost visualisation in Fig.~\ref{fig::QCDVis_main}.  Selected nodes, arcs, and corresponding contours are coloured green.  Nodes, arcs, and corresponding contours under the mouse cursor are highlighted in orange.
Following a similar approach as to Bajaj et al.~\cite{bajaj1997contour} the histogram presented alongside the isovalue selection slider shows the number of distinct objects in the scalar field across the isovalue range.   

In many \lqcd{} field data sets the regions of interest are the positive and negative extremes; however, the contour spectrum can also show interesting features in the region of percolation (the region around zero).  Signatures of (anti\nobreakdash-)instantons show up in the histogram as highly persistent objects in the positive or negative direction.  Other field variables, such as the Polyakov loop, usually show a measure of symmetry around zero.

\paragraph{Orthographic projections of surfaces (Fig.~\ref{fig::QCDVis_main}.5)}  We also present surfaces existing in the scalar field using orthographic projections from each 2D plane.  These have a fixed footprint with the scale adjusted according to the extent of axis, with the temporal extent of the data often different to that of the spacial axes.  This is because in \lqcd{} the temperature is given as an inverse function of the number of time steps.  In situations where the physicist prefers not to use these projections they can be hidden.%, allowing the user to focus on other aspects of the visualisation.
	
Orthographic projections can be a useful tool for physicists when tracking objects in specific fields, for instance \antiinstanton s in the topological charge density field, and is best suited to verifying and analysing phenomena identified by the cooling code.  As part of the cooling process the position of minima and maxima in the action and topological charge density fields is output by the program.  By identifying these objects using the orthogonal views their structure can visualised in various 3D configurations of the 4D hypervolume.  Occlusion from overlapping surfaces is rarely a problem in this view, as the objects of most interest appear in sparse areas of the scalar topology.

\paragraph{Flexible isosurface view (Fig.~\ref{fig::QCDVis_main}.6)}  The core visual interface is a 3D rendered view of the selected scalar field.  Use of the contour tree algorithm enables the transition from a basic isosurface to representation of connected contours in the data. This has two benefits; first, optimized visualisation techniques can be applied to the data.  Second, properties can be calculated on distinct contours as opposed to the level set as a whole, allowing the calculation of \emph{physical} measurements of objects existing in the data.

Each arc in the contour tree is a representative of a unique object, the mesh for which is generated using a modified form of the \emph{flexible isosurface} algorithm~\cite{carr2010flexible}. Each isosurface can be continuously deformed between upper and lower bounds of the contour tree arc.  This offers the possibility for parallel computation of complex contours.

\paragraph{Properties of topological objects}  The triangle meshes created for rendering also allow us to query properties of the objects that are beyond the statistical mechanics features usually available to physicists.  The simplest of these is evaluation of the surface area and volume of the mesh. 

To calculate more advanced properties we use the mathematical concept of moments~\textemdash{} using methods initially implemented by Zhang and Chin~\cite{zhang2001efficient}.  The zeroth moment represents the signed volume of a contour; the first moment, the centre of mass; and the second representing moments of inertia.  A pre-requisite for correct moment calculations is for a mesh to be closed.  An initial step to determine this is to calculate the Euler-Poincar\'{e} characteristic ($\chi = V - E + F$) of a mesh.  Following this approach it is possible that a contour which spans two or more boundaries can be incorrectly identified as an $n$-torus.  To counteract this, objects are also tested to see if vertices lie on the height field boundaries.  A more in-depth discussion of the use of moments in visualisation can be found in~\cite{schlemmer2007moment} and~\cite{bujack2014moment}. We calculate the centre of mass using the three first-order moments, however we only show centre of mass and moment of inertia for closed contours.

Centre of mass requires the calculation of three first-order moments by computing a weighted average in the $x, y$, and $z$ planes of each triangle.  This is scaled by its contributing volume, calculated by placing an additional vertex at the origin to make a tetrahedron.  The final centre of mass for the object is calculated as a positional vector that must be scaled by the entire objects volume.  We only show the centre of mass for closed contours in the main display by a marker.  The moment of inertia requires the computation of the eigenvectors of the $2^{nd}$-order tensor.  First-order mixed moments, which take into account averages on the plane, need to be calculated along with second-order moments.  Eigenvalues and eigenvectors of the inertia tensor are computed, normalised, and sorted in order of magnitude of their derived eigenvalue with the largest being the principle component axis.  Comparison of the relative magnitudes of the eigenvalues, and orientation of the corresponding eigenvectors with respect to the $t$-axis, reveals any tendency for instanton distortion, either oblate or prolate, as $\mu$ is increased.
		
\paragraph{Periodic boundary conditions}  A substantial challenge is posed by the periodic nature of the data being handled in this application.  To ensure translational invariance of quantum averages, \lqcd{} simulations are performed with periodic boundary conditions for the gauge variables in the time and three spatial dimensions.  Hence, the extreme vertices of each dimension of the data are considered as being neighbours with a cell existing between them, as shown by Fig.~\ref{fig:periodicBoundaries}.  For the purposes of generation this gives the illusion of an infinite universe but complicates computation of the contour tree.  The Reeb graph representation, however, does not suffer from this limitation.  We therefore supply the user with both representations to further support the analysis. In Figure~\ref{fig:periodicBoundaries} the comparison of the Reeb graph and contour tree for each view shows that the Reeb graph correctly sees the object as one distinct superarc (topological object).  However, in the contour tree the surfaces are represented by two separate superarcs.  Periodicity of data is a feature not limited to QCD datasets but also appears in other fields of science including computational chemistry~\cite{beketayev2014extracting} and computation biology~\cite{alharbi2016molpathfinder}.

%-------------------------------------------------------------------------

\section{Results}

\noindent
The QCDVis framework was developed following a nested model approach~\cite{Munzner2009NestedModel}.  Close collaboration with physicists was crucial throughout the four stages of characterisation, abstraction, and design.  The model was merged with a standard iterative design with each stage revised at the end of each cycle together with the domain experts.  The approach we adopted allowed the domain experts to gradually familiarise with the framework raising suggestions, questions, and proposal of new directions for development of novel or alternative features.  

%There were crucial stages during development \red{where the domain experts had eureka moments} and became aware of opportunities for data inspections made possible by our tool.  Two particular instances led to the formulation of two case studies to evaluate the effectiveness of the framework to support specific analytical tasks.  

In order to demonstrate possible uses of the QCDVis framework, we carried out user testing by implementing two case studies.  The two tasks were designed in conjunction with the domain experts to reflect an optimal work flow for identifying topological objects within existing \lqcd{} data sets.  Data sets are obtained from the UK based DiRAC facility which computes and maintains a number of \lqcd{} projects.  Data is pre-cooled using code supplied by Hands and Kenny~\cite{han2011a}~\textemdash{} this process is beyond the scope of this paper.

Here we review the feedback from the case studies, a more detailed written account of the studies can be found in the appendix.  A video is available at https://vimeo.com/205054908 that shows the system in operation.

\subsection{Domain expert feedback and observations}

\noindent
Feedback gathered during the hands-on case studies sessions was largely positive with the framework welcomed as a new way of viewing and understanding \lqcd{} data.  Domain experts were taken aback by the possibility of visualising instanton shapes and computing physical variables such as surface areas and volumes, features they had never been able to visually inspect and measure before.  Particularly in the case where an anomaly was detected the visual feedback compelled a new way of interaction with the data. 

New suggestions on how to enhance existing visualisations were also brought forward; for example, the ability to view ensemble average isovalue histograms in addition to configuration histograms on the global isovalue slider.  It was felt that this enhancement could further steer users towards interesting features in the data that were previously beyond the reach of physicists who were unable to calculate the topology of their data.

One domain expert was particularly interested in the possibility of using the program to view data from other simulations belonging to research projects he is currently involved. Their work focuses upon 2 colour configurations generated by the MILC collaboration using a variation of the cooling algorithm.  They believed that the tool represents an interesting method for visually examining the difference between differing cooling algorithms.  The existing modular design of QCDVis should allow the relevant file format to be incorporated into the program with minimal effort.

Additionally, we noticed a difference in the ability to make full use of the contour tree or Reeb graph between physicists that had been involved in the development of the framework from the very beginning. Experts with less prior experience of using the contour trees tended to focus more on the isosurface representations of the data.  However, as the interaction with the framework progressed, they felt that the contour tree and Reeb graph were offering interesting tools for obtaining an overview and navigating through the data.

%-------------------------------------------------------------------------
\section{Conclusions}
\noindent
We have presented a framework for visualisation and analysis of \lqcd{} data sets using topological features for segmentation.  The system is designed with domain scientists to ensure that the abilities suit the requirements of a typical user in the field.  The system is able to extract the topology of field variables and compute multiple object meshes using established algorithms in parallel on a single system.  The system already allows researchers to look at their data in new ways, encouraging them to think about properties of their data in ways previously beyond their reach.   

The ability to view data from multiple fields in parallel within the application allows the user to make observations about correlations between lattice fields.  This was found to be particularly useful when examining the plaquette fields and the topological charge density in parallel as the values in the plaquette fields contribute to topological charge density.  Viewing surfaces extracted from the data alongside the topological structure, as captured by the Reeb graph, also allows new understandings to be formed in less ordered fields such as the Polyakov loop.

At present the system has been used to view data from a specific set of 2 colour QCD experiments, which it is able to handle with a satisfactory computation time.  As a result feedback from the case studies, we have also had interest in using the tool on two colour simulations from the MILC QCD project.  These differ from the existing configurations in using a less discrete form of cooling.  We therefore envisage the use of the system to help physicists to evaluate the relative merits and side effects resultant from the choice of cooling algorithm.

However, the potential for larger configurations could require the application of distributed computing environments in order to and maintain interactive framework.  Therefore, further optimisations to the application could be incorporated, as discussed by Bremer et al.~\cite{bremer2011interactive}.

%-------------------------------------------------------------------------
\section{Future work}
\noindent
Feedback from physicists suggested an interest in the comparison of scalar fields of different variables.  This is demonstrated in the first case study by the close proximity of minima and maxima within the Wilson action and topological charge density fields.  As a result of insights gathered from the use of QCDVis we are in the process of an ongoing study into effects that varying chemical potential has on topological structures existing on the lattice.

\section{Acknowledgements}

\noindent
This work used the resources of the DiRAC Facility jointly funded by STFC, the Large Facilities Capital Fund of BIS and Swansea University, and the DEISA Consortium (www.deisa.eu), funded through the EU FP7 project RI- 222919, for support within the DEISA Extreme Computing Initiative.  The work was also partly funded by EPSRC project: EP/M008959/1.  The authors would like to thank Dave Greten for proof reading this document.

%Carr and Duke~\cite{Carr2013} consider the use of the Joint Contour Net (JCN) for the purpose of looking for relationships within multivariate data.  The method presented allows new visualisations and analysis technique to be employed for looking for patterns in the multivariate fields.  More recently, the problems of noise within scalar field data that made multivariate analysis difficult has been addressed by employing a topological simplification on the data ~\cite{huettenberger2014decomposition}.  Alternative approaches to the Joint Contour Net look for similarities within merge trees~\cite{beketayev2014measuring} and Reeb graphs~\cite{Bauer2014measuring} to identify patterns in complex data.
%
%The inclusion of one or more of the techniques described above should allow us to further exploit the use of topological visualisation in Lattice QCD.  The ability to examine the data using new techniques such as JCN may allow researchers to think about their data using a new approach.  Additionally, the ability to analyse the data for regions of topological similarity could present domain scientists with interesting new questions to ask of their data at the ensemble level.

%% References


\bibliographystyle{model3-num-names_nourl}
\bibliography{/home/dean/Documents/library.bib}

%% not want to use model3-num-names.bst.





% \end{thebibliography}

%-------------------------------------------------------------------------

%\ifthenelse{\boolean{APPENDIX}}
%{
\appendix
%-------------------------------------------------------------------------
\section{Extended domain background}
\label{sec::extended_domain_background}

%\subsection{Domain Specific Terminology}

%Here we define terms used by physicists to describe the hierarchical structure and operators existing used on their data.  These terms are used in the remainder of the paper.

\paragraph{Quantum Chromodynamics}

Quantum Chromodynamics (QCD) is the modern theory of the strong interaction between elementary particles called quarks and gluons, responsible for binding them into strongly-bound composites called \textit{hadrons}. The most familiar examples are the protons and neutrons found in atomic nuclei. It is an example of a relativistic quantum field theory (RQFT), meaning it is formulated in terms of field variables, generically denoted $\phi(\vec{x})$, living in a four dimensional space-time (note we often use $\vec{x}$ to stand for the set of four space-time coordinates $x,y,z,t$). It is convenient to work in Euclidean metric where time is considered as a dimension on the same footing as $x,y,z$; technically, this is done by analytic continuation from Minkowski space-time via $t\mapsto \imagi t$ a combination of Euclidean space and time into a four dimensional manifold.  In this Euclidean setting, any physical observable $O(\phi)$ in QCD can be accessed via its quantum expectation value

\begin{equation}
\langle O\rangle=\mathcal{Z}^{-1}\int\mathcal{D}\phi O(\phi)\exp(-S(\phi)),
\label{eq:FPI}
\end{equation}

where the \textit{action} $S=\int d^4\vec{x}\mathcal{L}(\phi,\partial\phi)$ is defined as the spatial integral of a Lagrangian density $\mathcal{L}$ which is local in the fields and their space-time derivatives, and $\mathcal{Z}$ is a normalisation factor such that $\langle 1\rangle\equiv1$, in other words, the expectation value of the unit operator is defined to be one. The symbol $\mathcal{D}\phi$ denotes a functional measure implying all possible field configurations $\phi(\vec{x})$ are integrated over. The expression (\ref{eq:FPI}) is known as a \textit{path integral}. 

\paragraph{Lattice Simulations}

\begin{figure}[!htb]
	\centering
	\begin{minipage}{.48\columnwidth}
		\includegraphics[width=\textwidth]{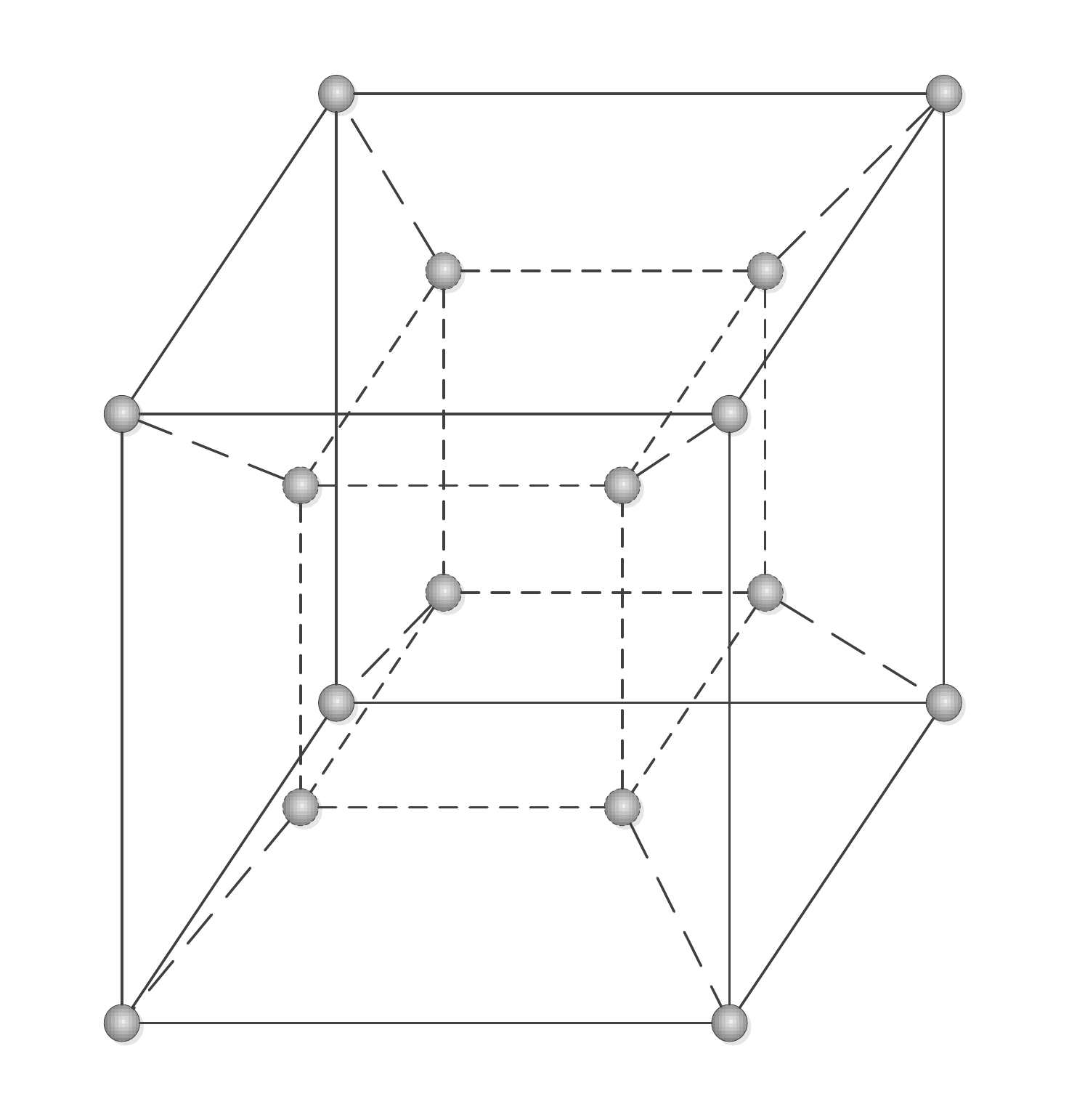}
		\caption{Arrangement of lattice vertices for one 4D cell (dashed lines represent $t$ dimension).}
		\label{fig::lattice}
	\end{minipage}
	\hfill
	\begin{minipage}{.48\columnwidth}
		\includegraphics[width=\textwidth]{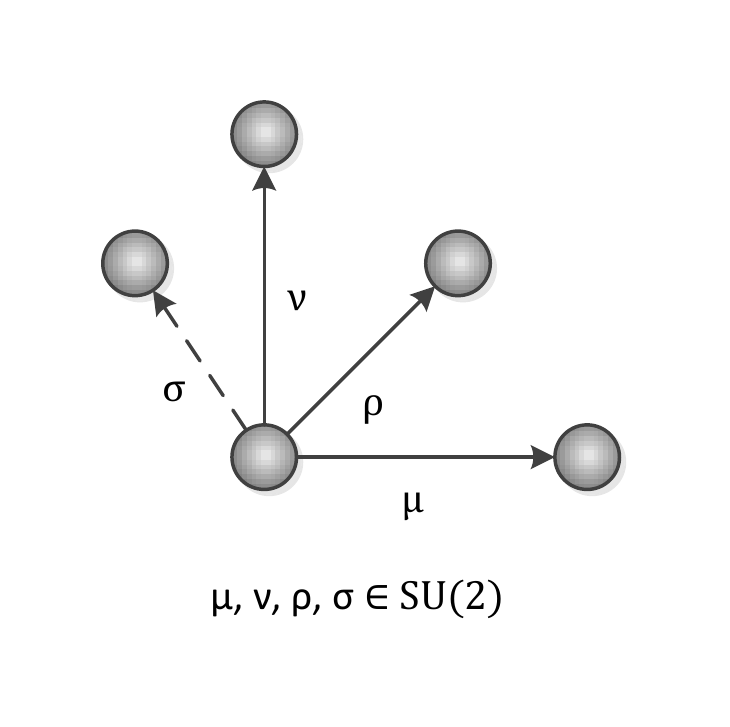}
		\caption{Placement of data on the 4D lattice edges ($\sigma$ is a link in $t$ dimension).}
		\label{fig::links}
	\end{minipage}
\end{figure}

\begin{figure}[!htb]
	\centering
	\includegraphics[width=\columnwidth]{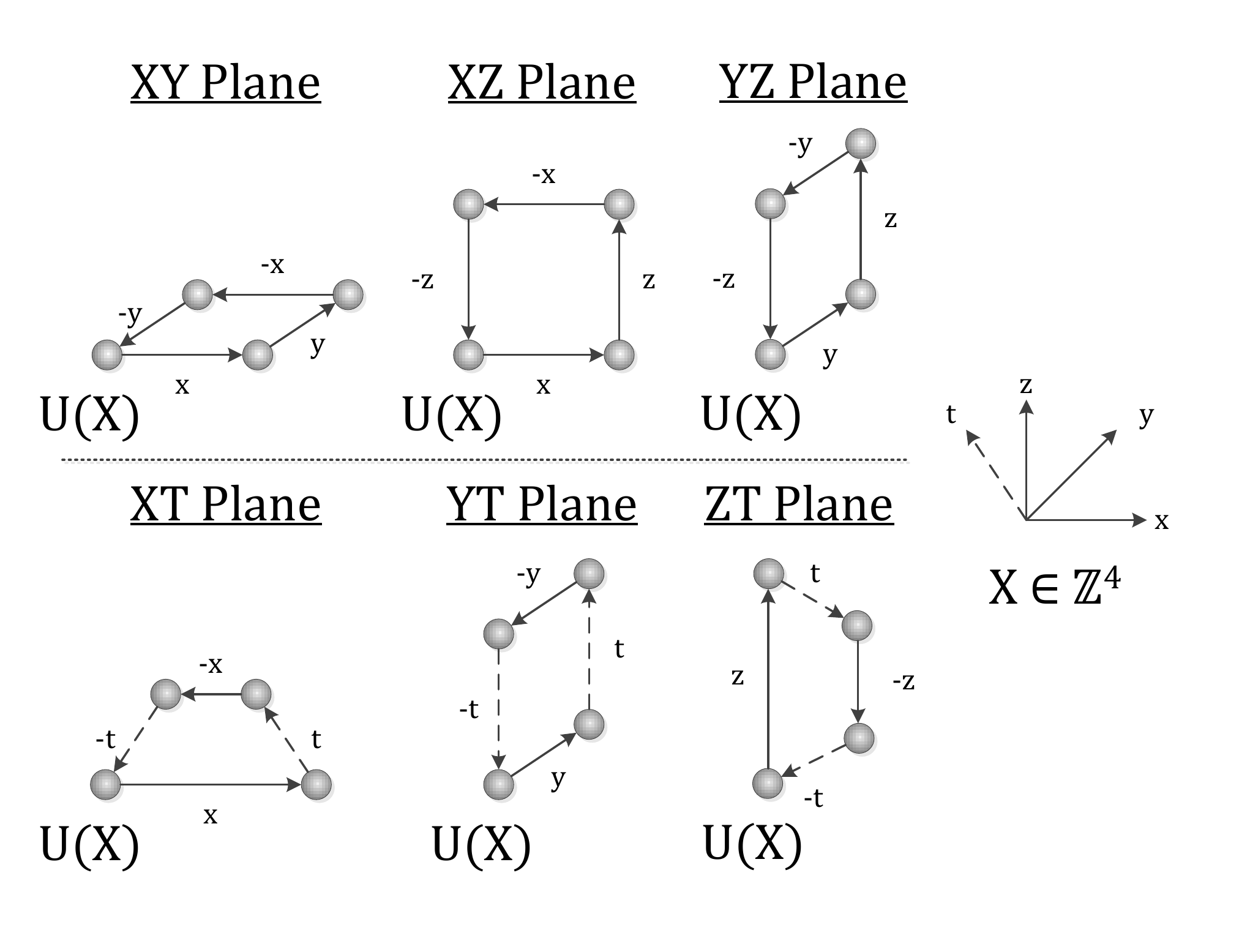}
	\caption{Top: the three space-like plaquettes on the lattice.  Bottom: the three time-like plaquettes on the lattice.}
	\label{fig::plaquettes}
\end{figure}

The most systematic way of calculating the strong interactions of QCD is a computational approach known as \textit{lattice gauge theory} or \textit{lattice QCD}. Space-time is discretised so that field variables are formulated on a four dimensional hypercubic lattice (Fig.~\ref{fig::lattice}). Gluon fields are described by link variables $U_\mu(\vec{x})$ emerging from the site $\vec{x}$ (specified by 4 integers) in one of four directions $\mu$ (Fig.~\ref{fig::links}). $U_\mu(\vec{x})$ is an element of the Lie group $\su(N_c)$, where the number of colours $N_c=3$ for QCD, and is thus represented by an $N_c\times N_c$ unitary matrix of unit determinant. The $U_\mu$ are oriented variables, namely $U_{-\mu}(\vec{x}+\hat\mu)\equiv U_\mu^\dagger(\vec{x})$, where the dagger denotes the adjoint of the link variable, ie. the complex conjugate of its transpose. The action $S$ is given in terms of a product of link variables around an elementary square known as a \textit{plaquette} (see Fig.~\ref{fig::plaquettes})

\begin{eqnarray}
S(U)&=&-{\beta\over{N_c}}
\sum_{\mu<\nu}\sum_x\mbox{Re}\mbox{tr}U_{\mu\nu}(\vec{x})\equiv-\beta\sum_x s(\vec{x})\nonumber\\
\mbox{with}\;\;\;\;
U_{\mu\nu}(\vec{x})&=&U_\mu(\vec{x})U_\nu(\vec{x}+\hat\mu)U_\mu^\dagger(\vec{x}+\hat\nu)U_\nu^\dagger(\vec{x}).
\label{eq:Wilson}
\end{eqnarray}

%-------------------------------------------------------------------------
%\begin{figure}[!htb]
%	\centering
%	\includegraphics[width=\columnwidth]{baryon.pdf}
%	\caption{Colour neutrality in a tri-quark system (Baryon).}
%	\label{fig::colour_confinement}
%\end{figure}

where $\hat\mu$ is a one-link lattice translation vector along the $\mu$ direction.  Lattice QCD provides a precise definition of the path integral measure $\mathcal{D}\phi$ as the product of the $\su(N_c)$ Haar measure $dU$ over each link in the lattice. The multidimensional integral (\ref{eq:FPI}) is then estimated by a Monte Carlo importance sampling based on a Markov chain of representative configurations $\{U_\mu\}$ generated with probability $\propto e^{-S(U)}$. The parameter $\beta$ in (\ref{eq:Wilson}) determines the strength of the interaction between gluons; large $\beta$ drives the plaquette variables to the vicinity of the unit matrix $\mathbbm{1}_{N_c\times N_c}$ resulting in weakly fluctuating fields, and \textit{vice versa}. Such calculations, known (even if imprecisely) as \textit{simulations}, require state-of-the-art high performance computing resources, particularly if the influence of quark degrees of freedom is included.

%-------------------------------------------------------------------------

\paragraph{Instantons and Topological Fluctuations}

The non-linear field equations of QCD admit solutions with interesting
topological 
properties. Consider a solution whose action density $s(\vec{x})$  is localised near a particular
space-time point, and becomes negligible on a hypersphere
$S^3$ of large radius $r$ centred at this point. If we denote the corresponding link
fields in this limit by $U_\mu=\exp(i\omega\partial_\mu\omega^{-1})$ where
$\omega(\theta_1,\theta_2,\theta_3)\in$ $\su(N_c)$ and the $\theta_i$ parametrise
$S^3$, then the quantity
\begin{equation}
Q=-{1\over{24\pi^2}}\int d\theta_1d\theta_2d\theta_3\mbox{tr}\varepsilon_{ijk}
\omega\partial_i\omega^{-1}
\omega\partial_j\omega^{-1}
\omega\partial_k\omega^{-1}\in\mathbb{Z}.
\end{equation}
is quantised and is known as the \textit{topological charge} of the configuration. $Q$ is conserved, in the sense that in order to change its value the fields must be distorted through a region of configuration space where the action $S$ becomes very large. $Q$ can be expressed as the space-time integral over a density $q(\vec{x})$ which is a local function of the gluon fields. For smooth fields $q(\vec{x})$ resides on solutions of the field equations known as \textit{instantons} ($Q=+1$) or \textit{anti-instantons} ($Q=-1$).  Instantons are hyper-spherical with a spatial extent parametrised by a scale $\rho$, and action $S_I=4\pi^2\beta/N_c$.

For the rapidly fluctuating fields characteristic of the strongly interacting QCD vacuum, the density, size and shape of objects carrying $q(\vec{x})\not=0$ remains an open issue where modern visualisation techniques can help. In lattice QCD topological charge density can be represented by 

\begin{equation}
q_L(\vec{x})={1\over{32\pi^2}}\varepsilon_{\mu\nu\lambda\kappa}\mbox{tr}U_{\mu\nu}(\vec{x})U_{\lambda\kappa}(\vec{x}).
\label{eq:qlat}
\end{equation}
However, because the lattice formulation violates some of the assumptions underlying the notion of topological charge, the $Q$ resulting from integration of (\ref{eq:qlat}) is susceptible to short wavelength field fluctuations, and in general is not integer-valued. In order to unmask fluctuations of a genuinely topological nature a process known as \textit{cooling} is adopted, in which small local changes to the $U_\mu$ are made systematically to allow the system to relax into a local minimum of the action. Structures resembling instantons may then be identified~\cite{Teper:1985rb}. It is reasonable to expect the total $Q$ for a configuration to be relatively insensitive to the details of this process, but things are not so clear for more intricate properties such as size, shape and separation.  For instance, under cooling a nearby $I$ -$\bar I$ pair approach each other and then mutually annihilate once their separation is of the order of the lattice spacing.

\section{Case studies}
\subsection{Case study: Locate and visually inspect an instanton within a configuration}

\noindent
This task demonstrates the core purpose of QCDVis by enabling visual inspection of lattice observables.  A $16^3 \times 32$ ensemble from the Edinburgh DiRAC server is provided for the task with $\mu$ set to $0.7$.  The ensemble has been cooled for $30$ iterations and is known to contain (anti-)instantons; the task is to locate the (anti-)instantons and view them as objects in the topological charge density and space-like/time-like plaquette fields.  Existing methods inherited from statistical physics are able to identify the location of minima and maxima within the Wilson energy and topological charge density fields.  However, they are unable to reveal much about the structure of these observables or the surrounding features within the scalar fields at these points.

\begin{figure}[!htb]
	\centering
	\begin{minipage}{.48\textwidth}
		\includegraphics[width=\textwidth]{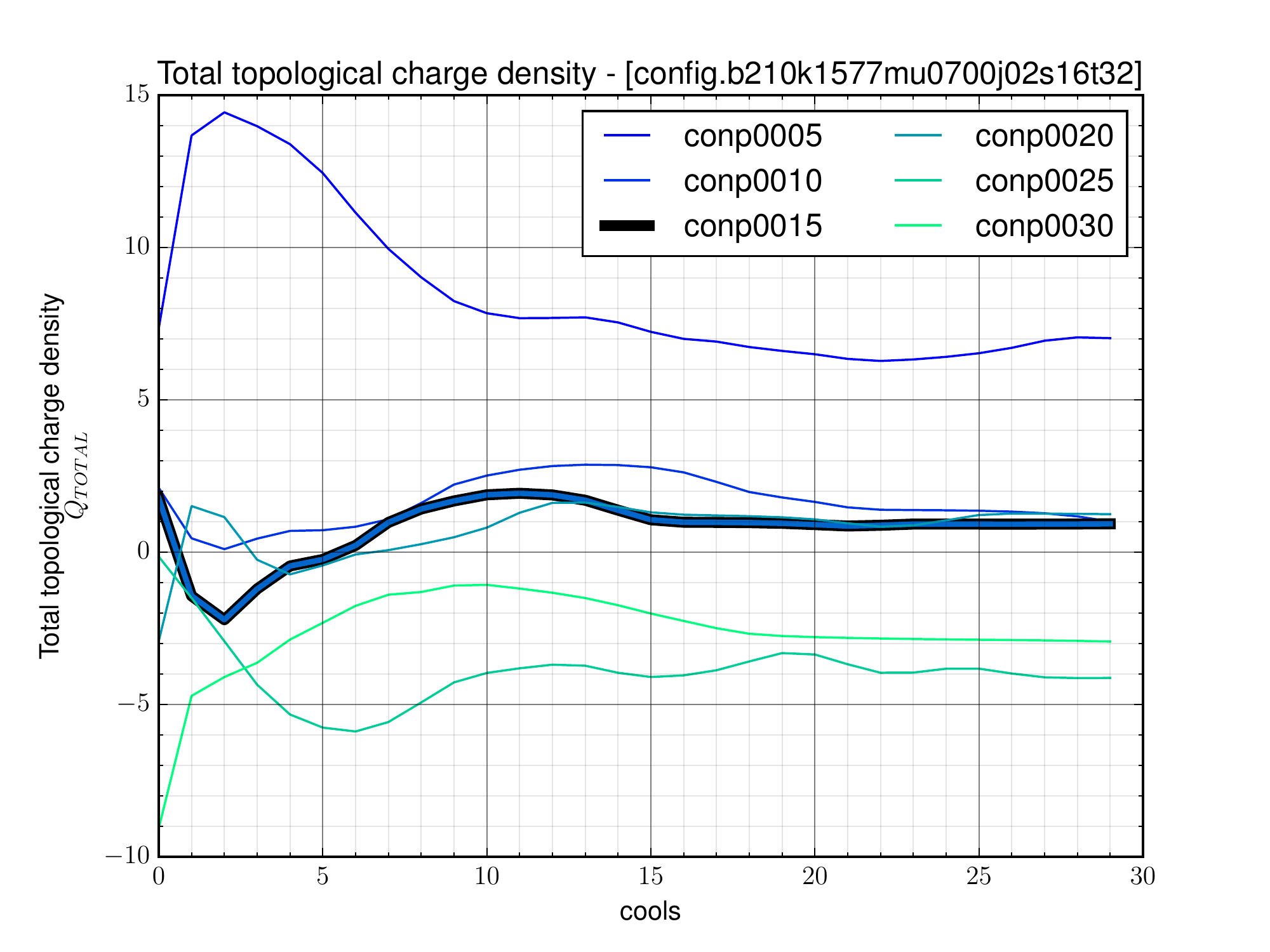}
		\caption{A flattening in the topological charge density indicates a region of stability that can be cross referenced against the peak action graph to select interesting levels of cooling for study.} 
		\label{fig::case_1_topological_charge_density}
	\end{minipage}
	\hfill
	\begin{minipage}{.48\textwidth}
		\includegraphics[width=\textwidth]{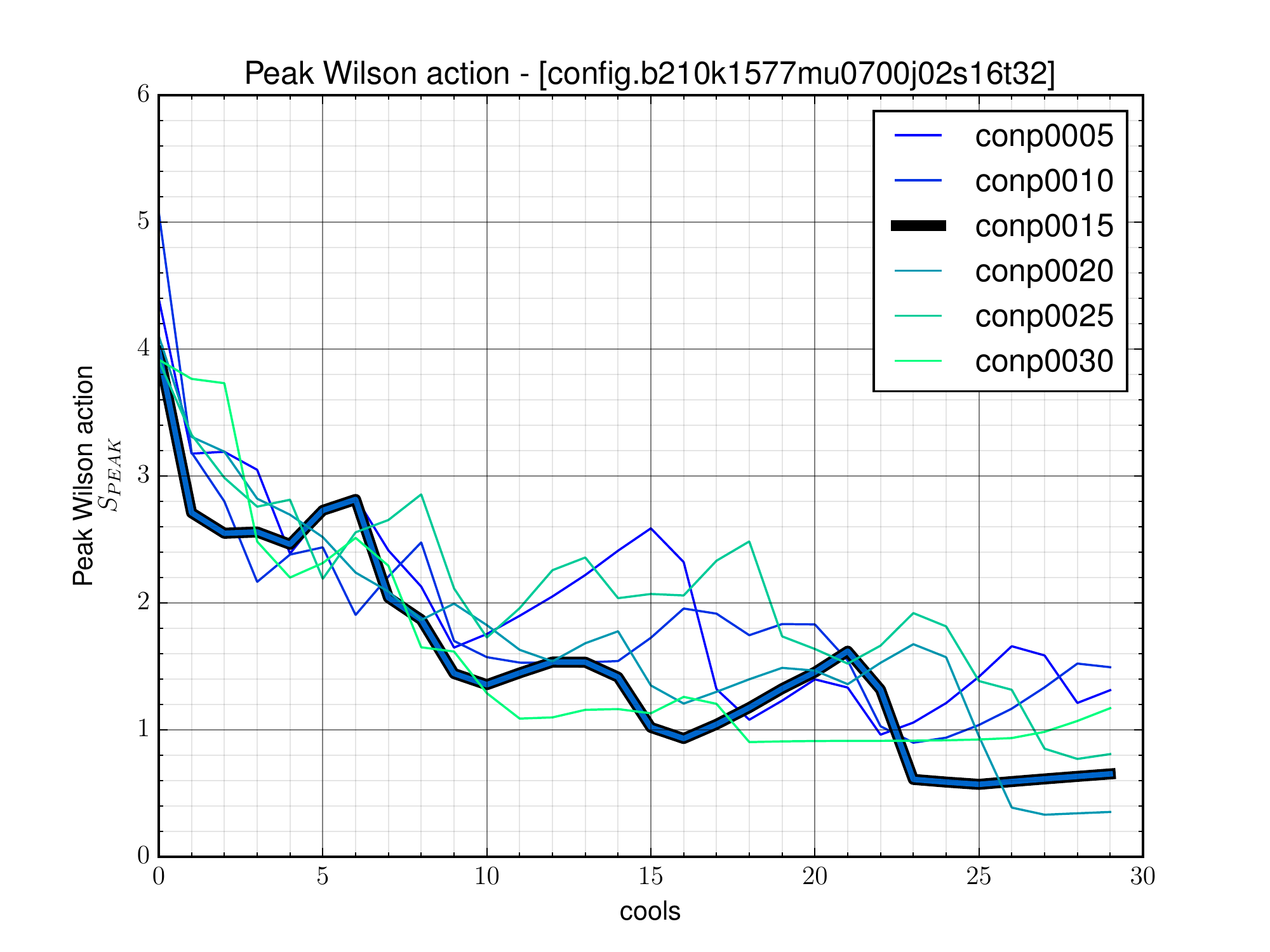}
		\caption{Spikes in the action graph indicate the spreading and smoothing of an instanton, followed by a merging of two or more distinct objects.}
		\label{fig::case_1_action}
	\end{minipage}
\end{figure}

Upon selecting an ensemble, an established first step for QCD researchers is to plot the action and topological charge density quantities for each configuration in the ensemble across the cooling domain.  This allows them to select a level of cooling at which they would typically conduct a study using ensemble data.  This is achieved by looking for signatures in the plots; of greatest interest is that of topological charge density, where flat areas of the graph indicate a stabilisation of the number of topological objects on the lattice (Fig.~\ref{fig::case_1_topological_charge_density}).  Cross-consulting the action max graph (Fig.~\ref{fig::case_1_action}) makes it possible to observe that the topological charge objects gradually increase in size within the region of stability, before merging at around 21 cools.  QCDVis supports physicists in this task by automating the task of generating each graph.

\begin{figure}
	\centering
	\includegraphics[width=0.45\textwidth]{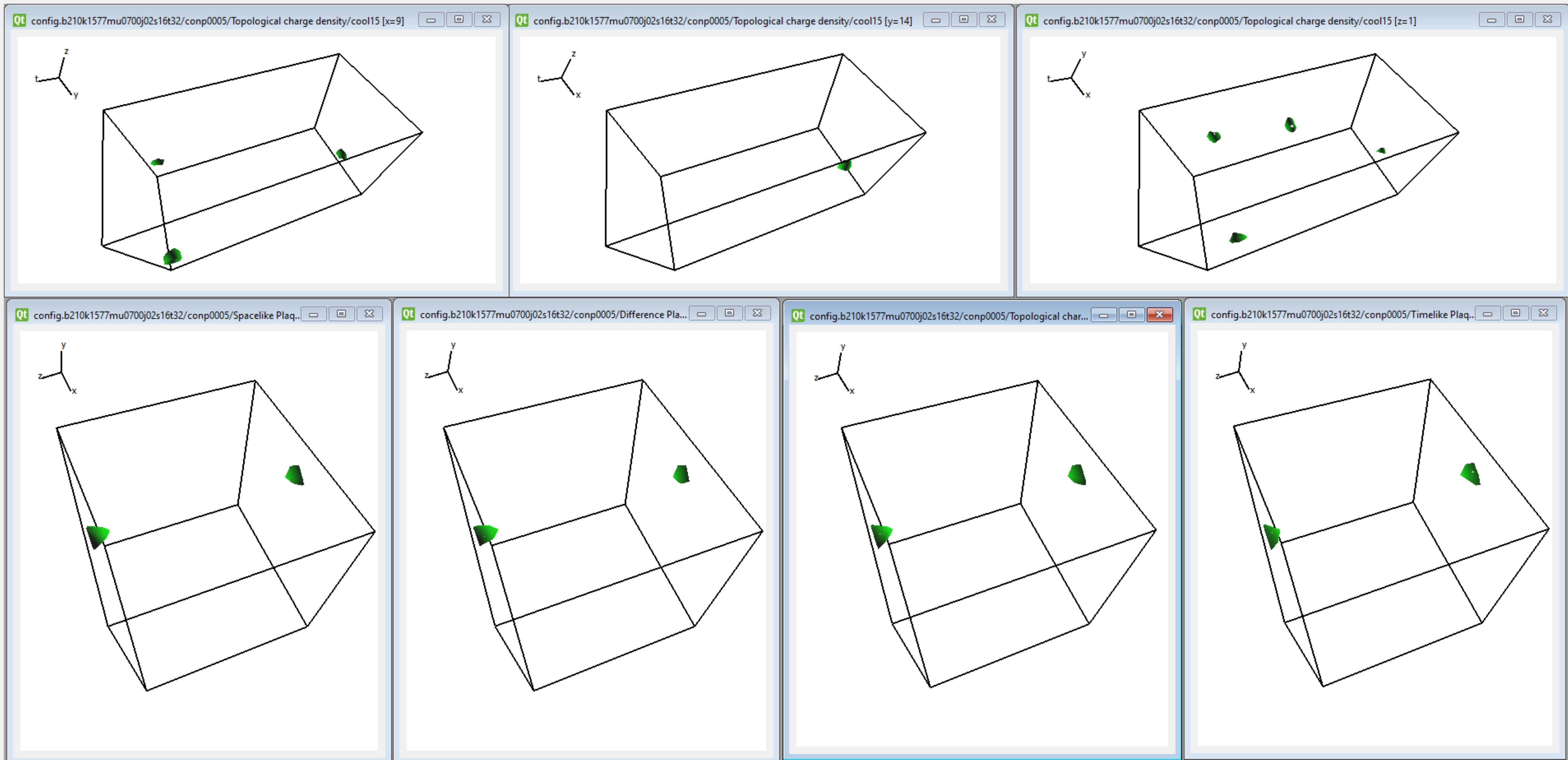}
	\caption{Top: user examining the position of Topological
		Charge Density minima (anti-instanton) in the $xyt$, $xzt$ and
		$yzt$ sub-volumes. Bottom: comparing the location of peak
		Wilson action and Topological Change Density minima in
		$xyz$ sub-volumes.}
	\label{fig::case1_main}
\end{figure}

To conduct a study on the physical form of the objects in the data, a configuration with a pronounced flat region was chosen.  Using the \textbf{ensemble module} (Sec.~\ref{sec::ensemble_module}) participants were able to select 3 levels of cooling for the desired configuration.  Field variables, in both the topological charge density and action fields, are important in this case study.  To understand the structure of the data, the researchers chose to represent the action density in the time-like and space-like configurations.  This enables them to look for obvious similarities in both variations, along with in the topological charge density field.

After computing desired field variables, focus switches to the \textbf{visualisation module} (see Sec.~\ref{sec::visualisation_module}).  Each field variable is loaded concurrently into the tool and the hyper-volume data cut into 3D configurations.  Feedback about the location of minima and maxima steers the researchers toward areas of interest for each field variable, observable in the contour tree.  Researchers tended to favour the 3D volumes split by time to locate variables in each field.  The global isovalue slider allows them to compare the relative similarities in each field in parallel.  Additionally, researchers study the location and structure of objects in the time-like 3D volumes ($xyt$, $yzt$, $xzt$) to validate that the objects coincide in all view configurations (Fig.~\ref{fig::case1_main}).
\subsection{Case study: View an irregularity within a cooled configuration.}

\begin{figure*}[th]
	\centering
	\begin{minipage}{.24\textwidth}
		\includegraphics[width=\textwidth]{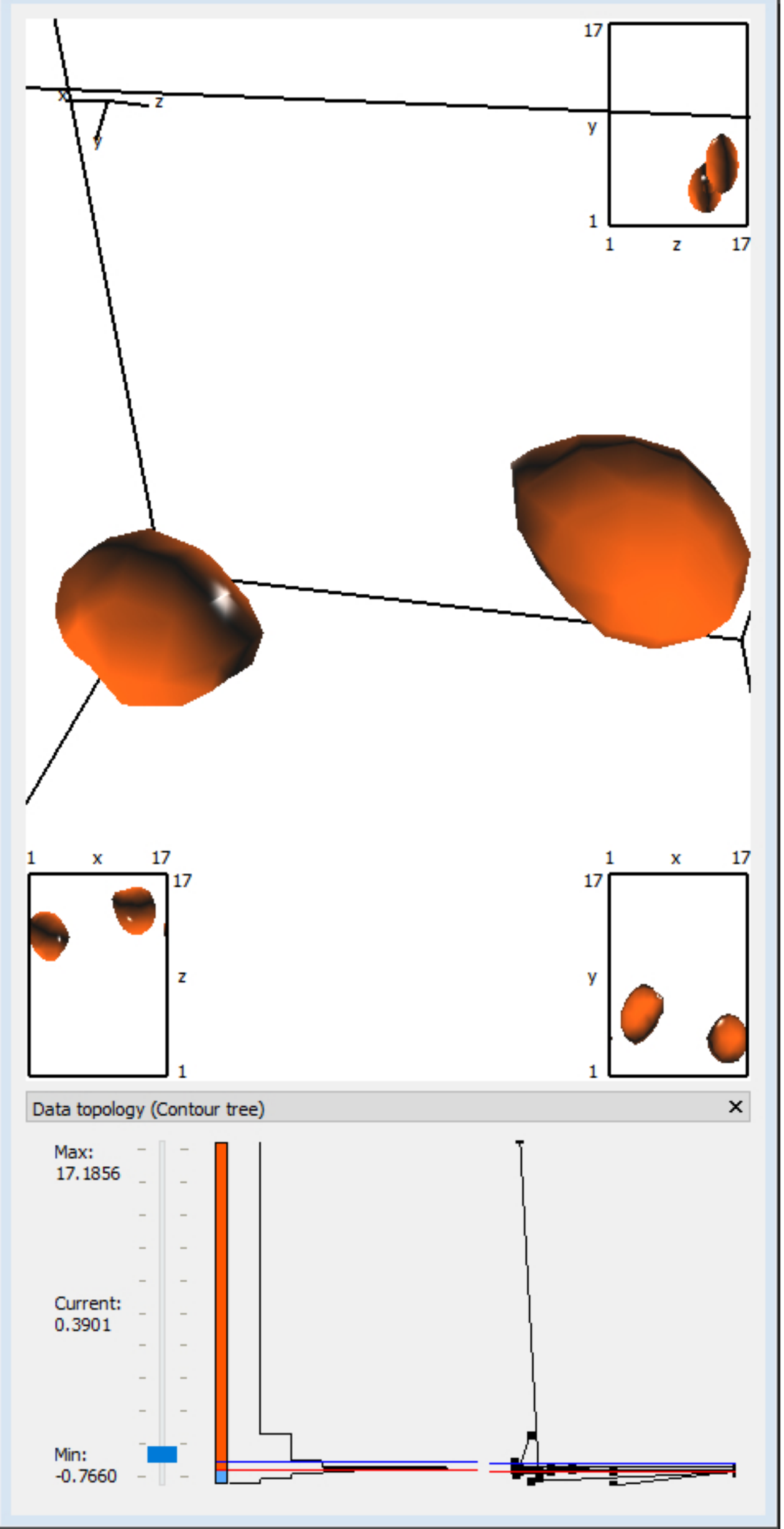}
		\caption*{Slice 35}
		\label{fig:Slice35}
	\end{minipage}
	\hfill
	\begin{minipage}{.24\textwidth}
		\includegraphics[width=\textwidth]{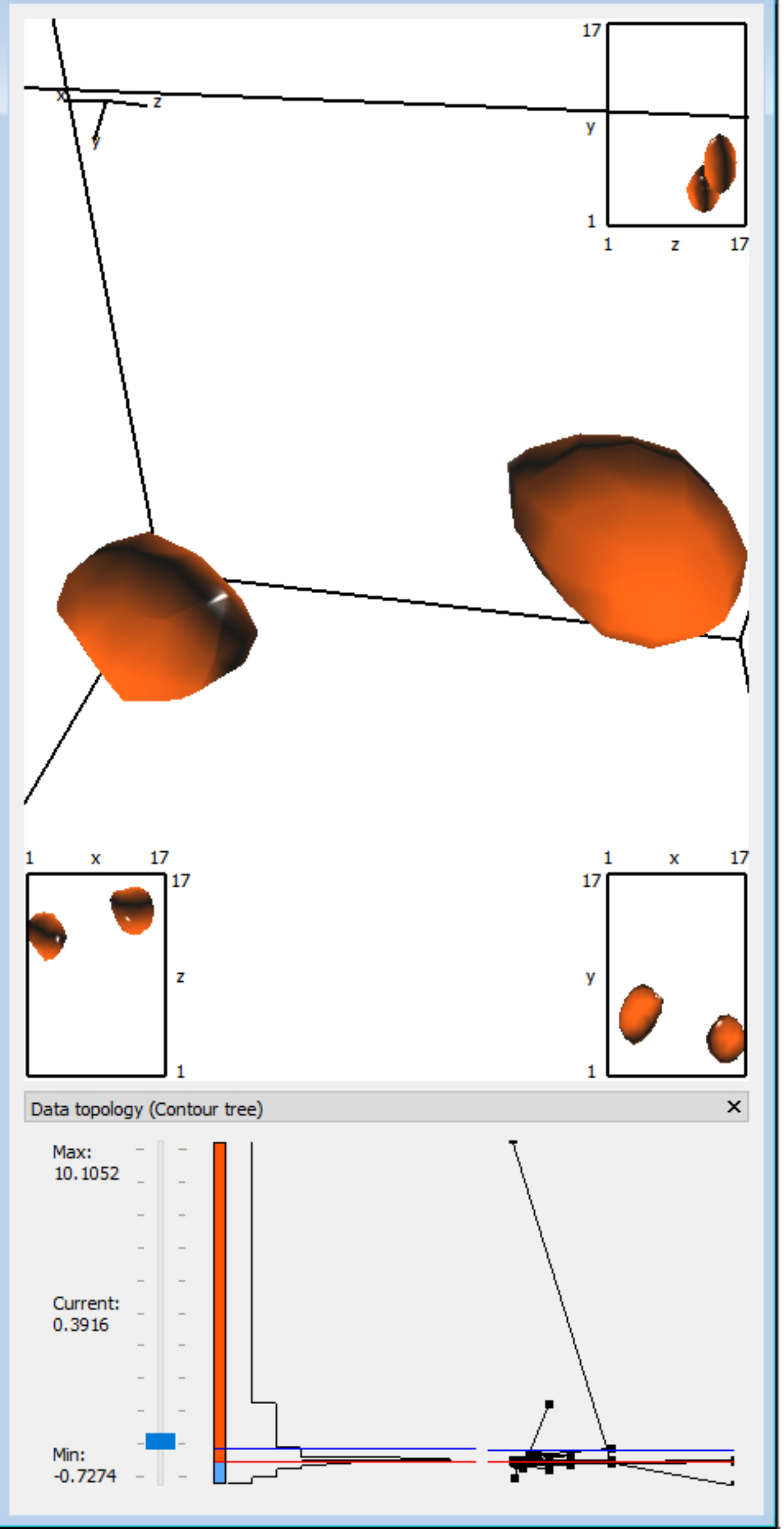}
		\caption*{Slice 36}
		\label{fig:Slice36}
	\end{minipage}
	\hfill
	\begin{minipage}{.24\textwidth}
		\includegraphics[width=\textwidth]{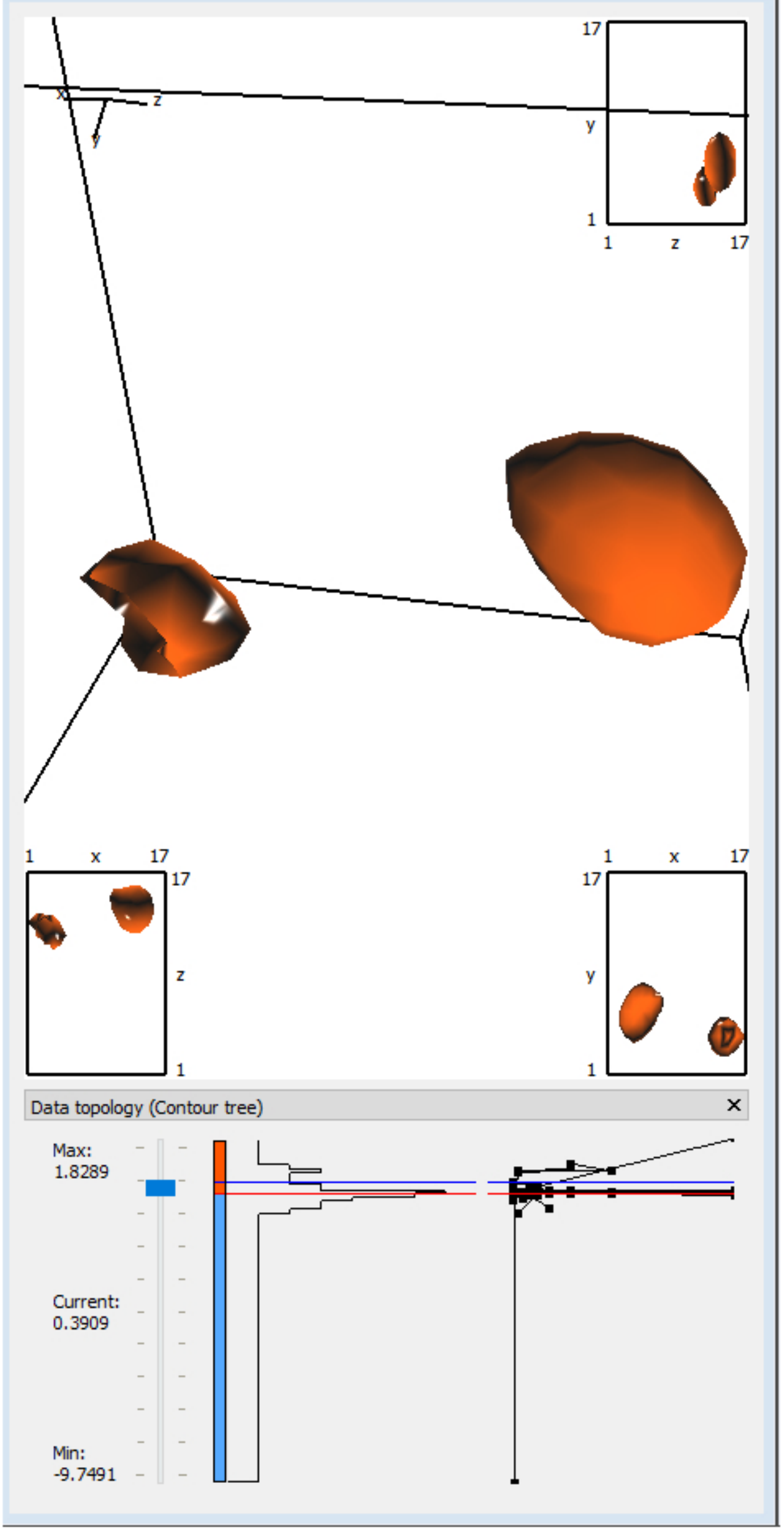}
		\caption*{Slice 37}
		\label{fig:Slice37}
	\end{minipage}
	\hfill
	\begin{minipage}{.24\textwidth}
		\includegraphics[width=\textwidth]{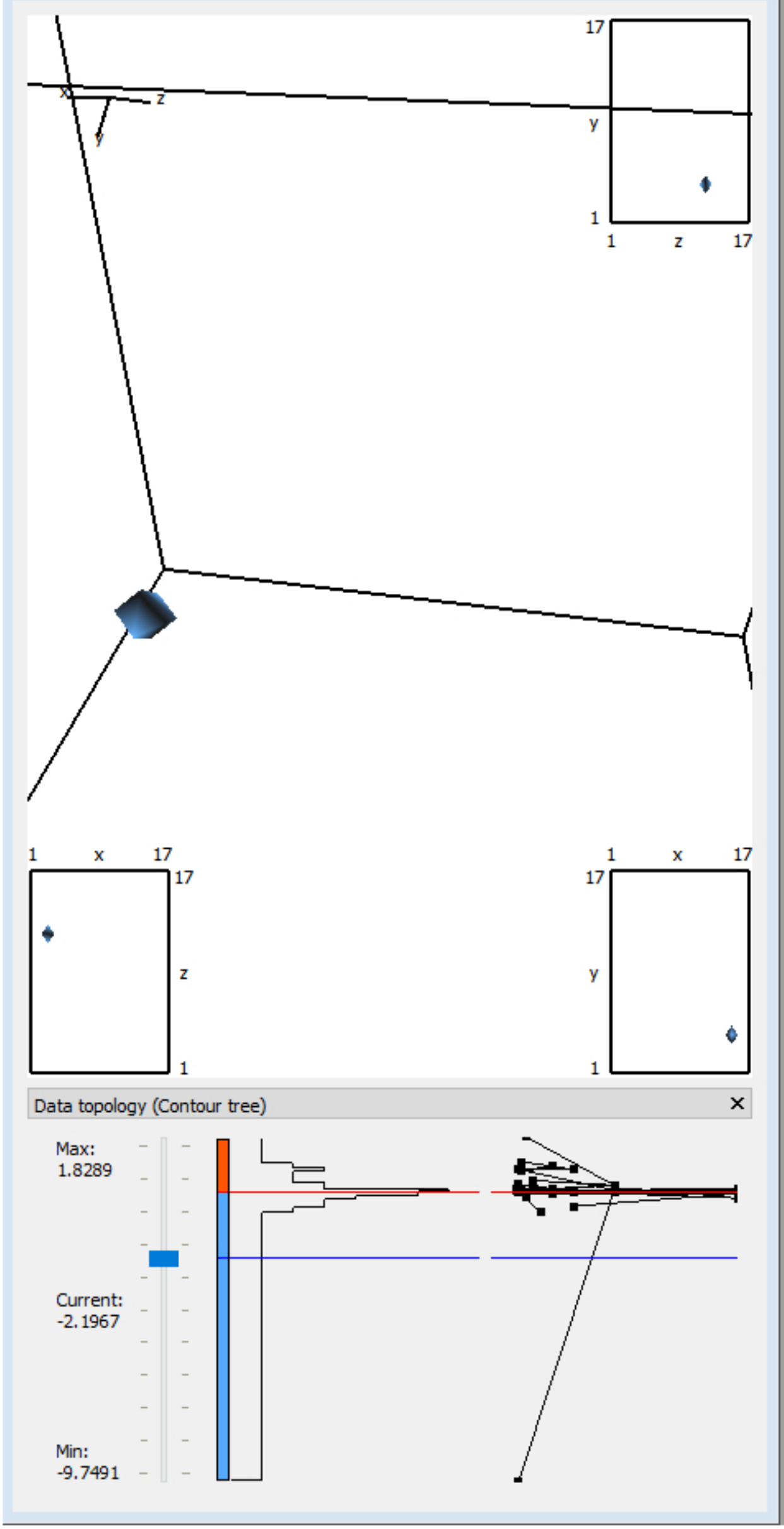}
		\caption*{Slice 37, anomaly.}
		\label{fig:Slice37a}
	\end{minipage}
	\caption{Cooling slices 35, 36 and 37, shown at the same isovalue.  At cooling step 37 the large object on the left starts to deform and disappears, being replaced by a global minima (right image).  This phenomena can also be observed in the contour trees for each cooling phase.}
	\label{fig::case2_main}
\end{figure*}

\noindent
The task illustrates how QCDVis is used to view and understand the cooling process in greater detail.  For this task a $16^3 \times 32$ ensemble from the Edinburgh DiRAC server is provided; however, in this ensemble $\mu = 0.45$.  The data has been cooled for $100$ iterations to try to emphasise the phenomena of observables dropping through the lattice.  An unexpected irregularity is found within one of the configurations, where a positive maxima becomes a negative minima in successive cooling cycles.  The task intends to visually inspect this phenomena and assign reason to it.

Whilst 100 cools represents a vast over exaggeration of how cooling would be applied in a real study, it promotes the possibility of witnessing an instanton \emph{falling through the lattice}.  This occurs when an instanton contracts to exist at a single point, whereas typically, instantons exist across multiple neighbouring sites.  In a typical QCD study this can only be detected using plots of topological charge density and action by looking for known signatures.

%\begin{table*}[!th]
%	
%	\begin{tabular}{|c|c|}
%		\hline
%		\textbf{Case 2} & \textbf{Case 2  - Close up} \\
%
%		&\includegraphics[width=0.30\textwidth]{case2_ffdual_tot}
%		&\includegraphics[width=0.30\textwidth]{case2_ffdual_tot_zoomed}\\
%		\emph{Total topological charge} & \emph{Total topological charge - close up} \\
%		\hline
%		\includegraphics[width=0.30\textwidth]{case2_action_max}&  \includegraphics[width=0.30\textwidth]{case2_action_max_zoomed}\\
%		\emph{Peak action} & \emph{Peak action - close up} \\
%		\hline
%	\end{tabular}
%	Case 2 - Singularity: close ups of Case 2 highlighting the signature of an instanton \emph{falling through the lattice}.  When choosing a configuration and level of cooling for evaluation, physicists will look for a flattening of the topological charge density. An instanton falling through the lattice is characterised by the relatively high gradient drop in the graph. }
%	\label{tab:Cases}
%\end{table*}

\begin{figure}[!htb]
	\centering
	\includegraphics[width=.48\textwidth]{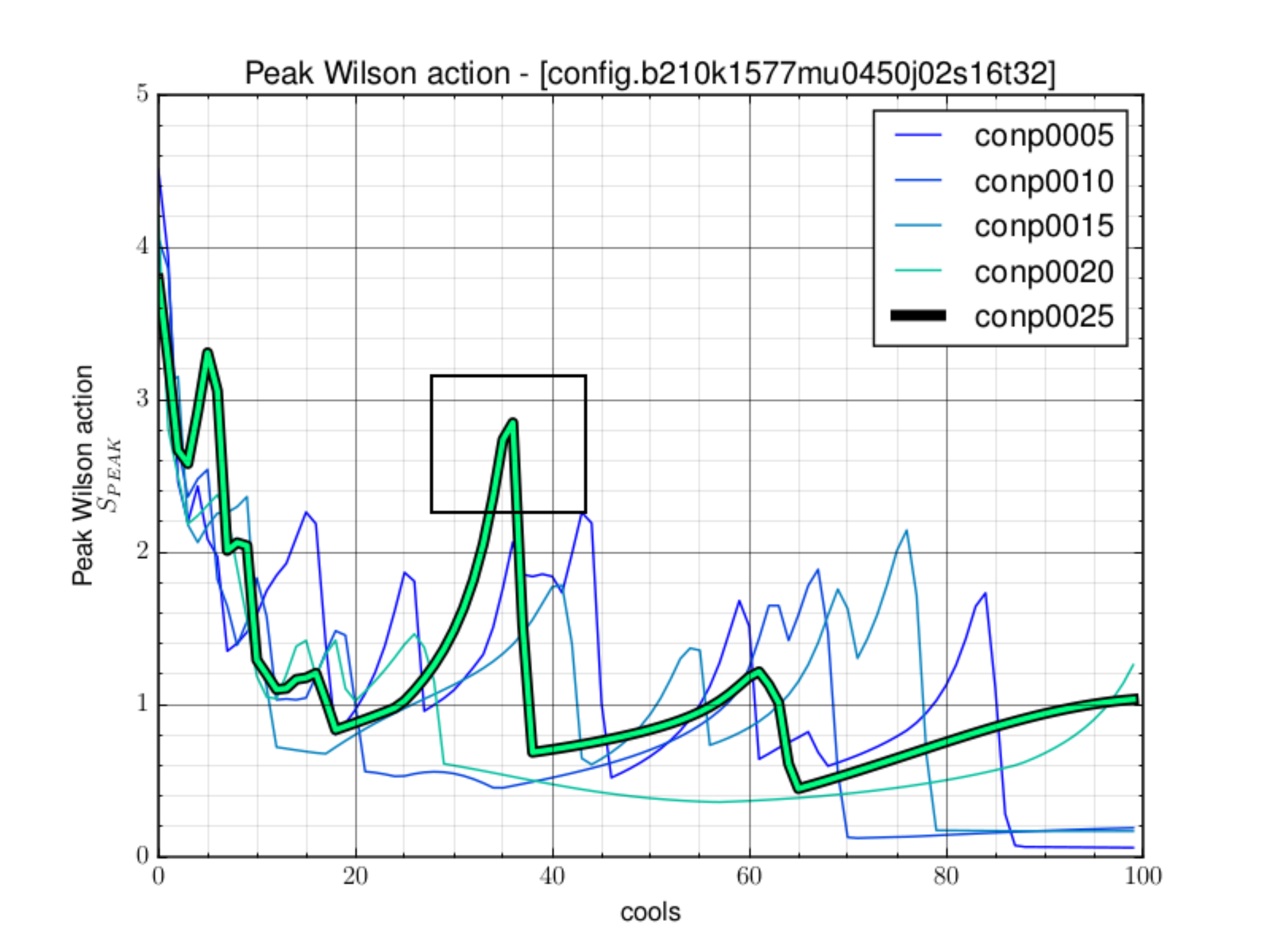}
	\includegraphics[width=.48\textwidth]{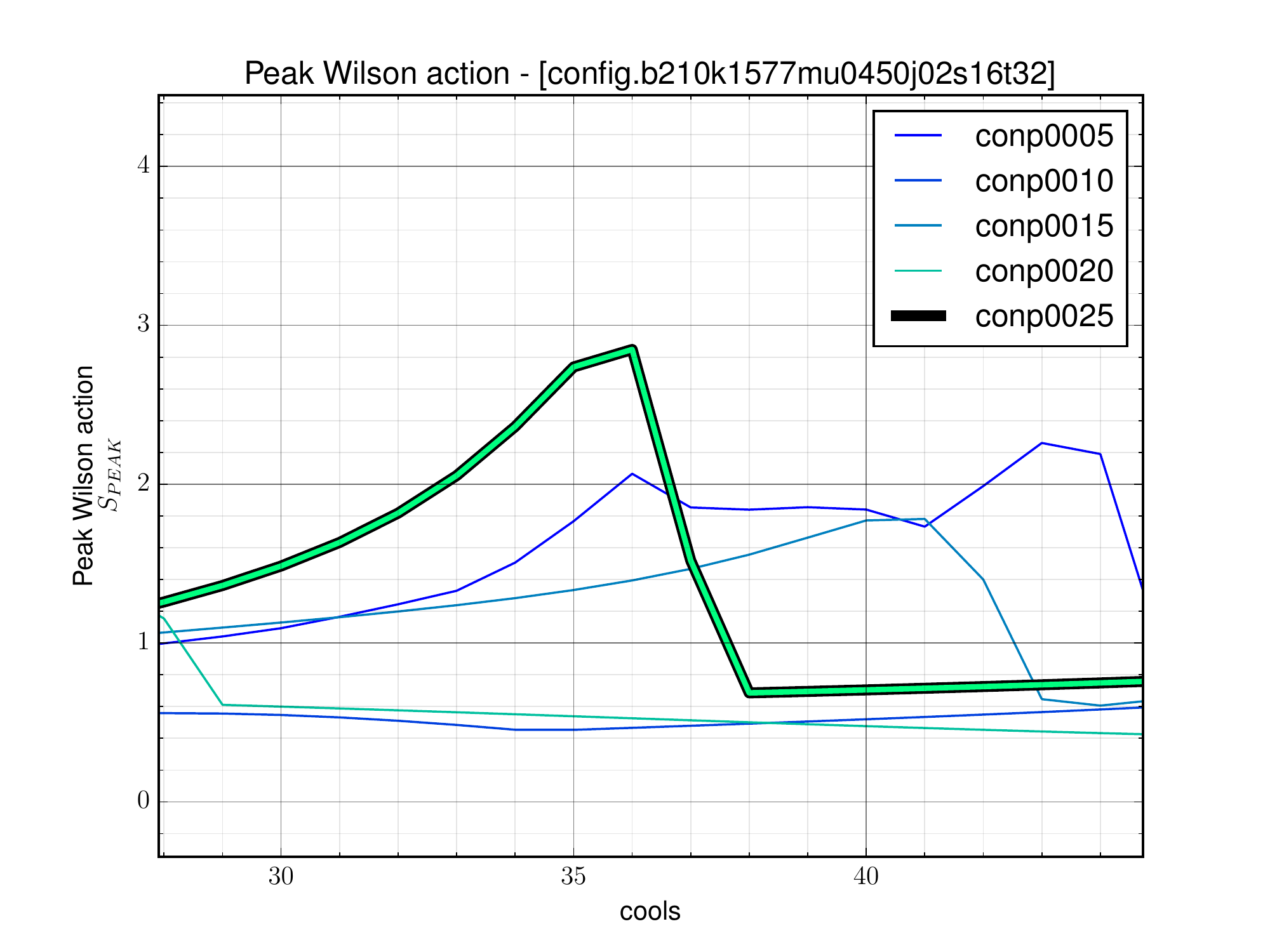}
	\caption{An instanton \emph{falling through the lattice} is characterised by the relatively high gradient drop in the peak action graph, this can be cross referenced with the topological charge density graph.}
	\label{fig::case_2_action}
\end{figure}

\begin{figure}[!htb]
	\centering
	\includegraphics[width=.48\textwidth]{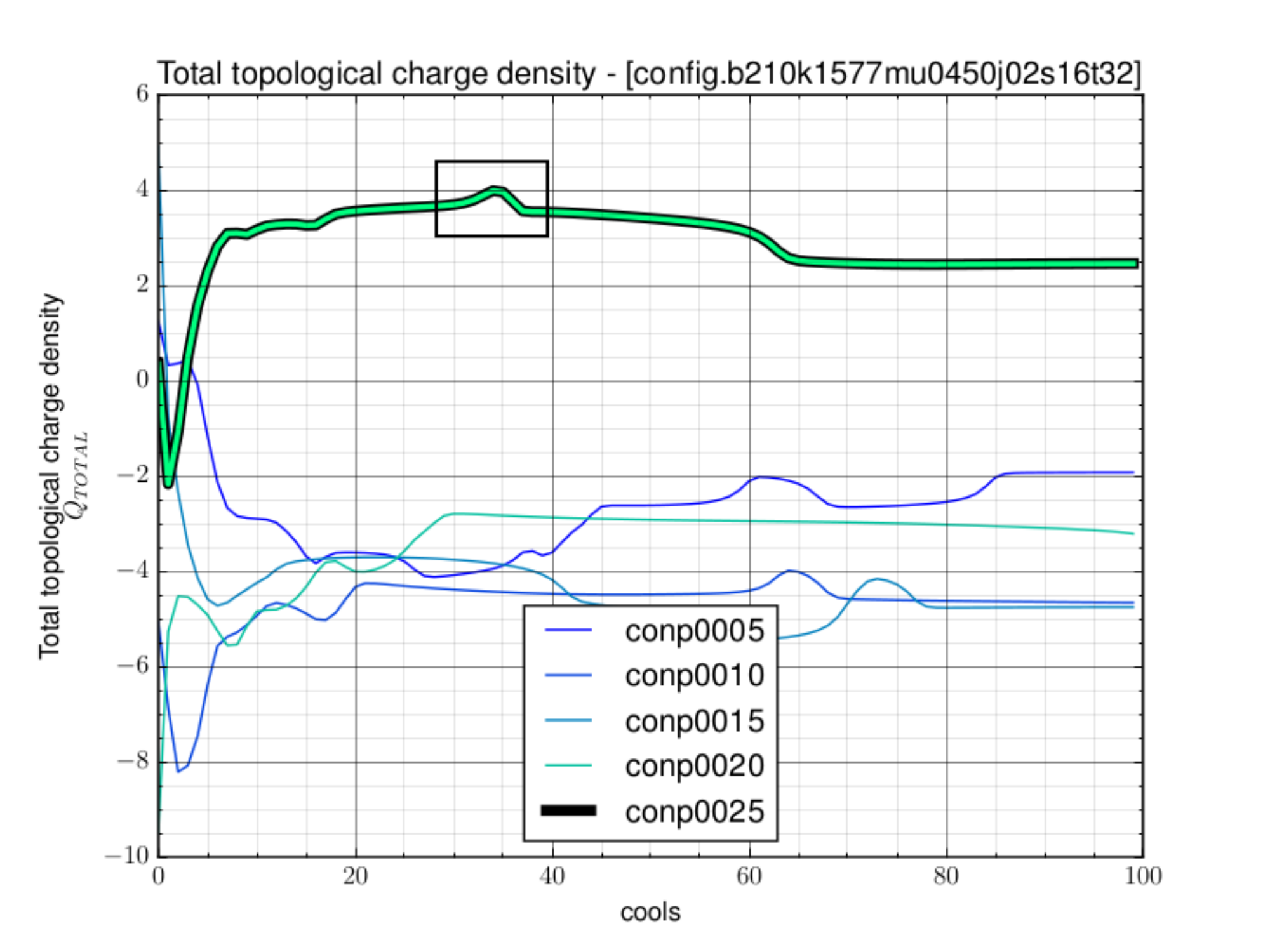}
	\includegraphics[width=.48\textwidth]{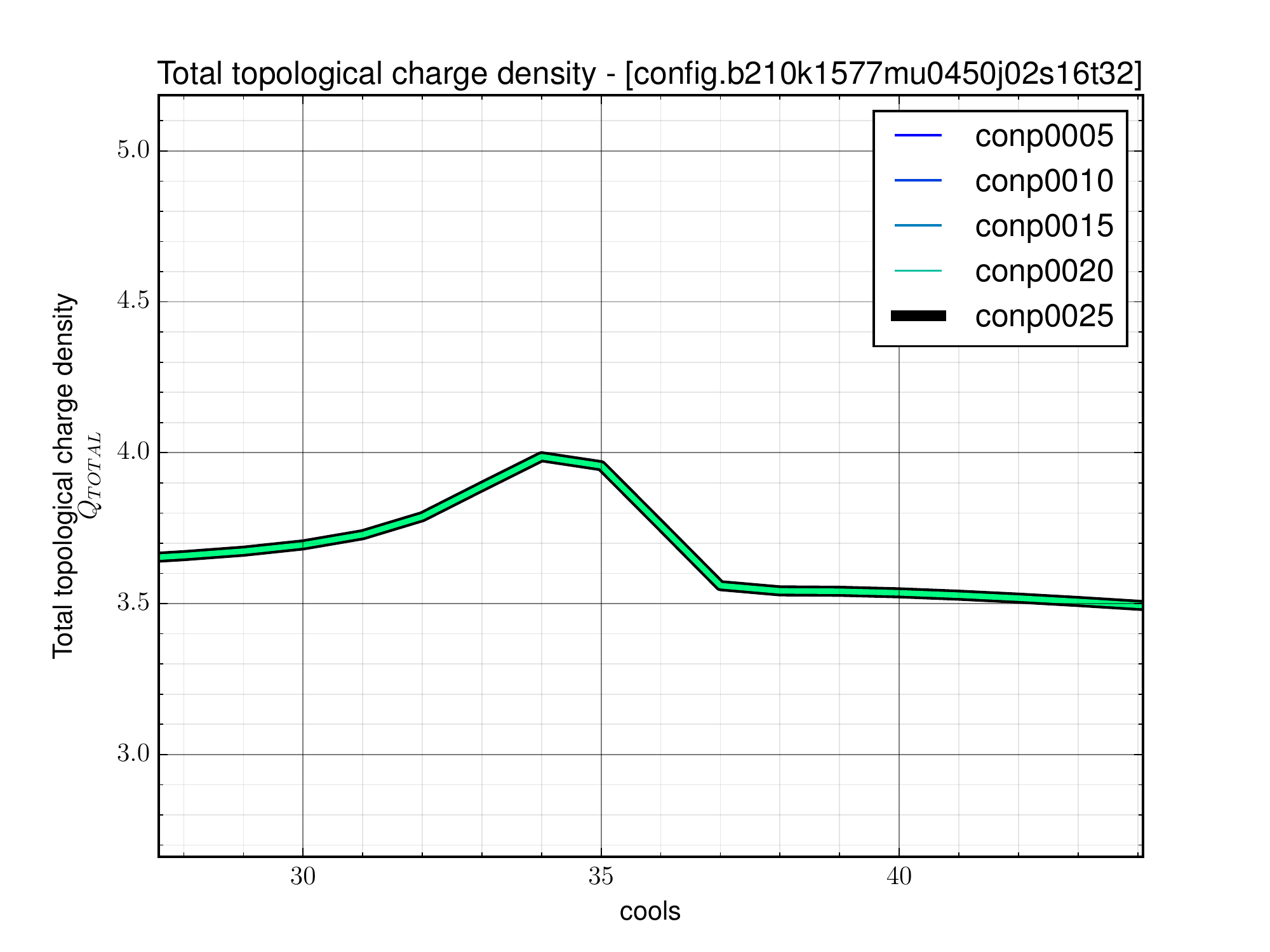}
	\caption{Physicists look for a flattening of the topological charge density indicating stability.  A short lived peak in the graph can characterise an instanton \emph{falling through the lattice} or an instanton-anti-instanton annihilation.}
	\label{fig::case_2_topological_charge_density}
\end{figure}

Initially upon loading a configuration the users consult the topological charge density and action max graphs.  However, in this scenario the focus is a gradual increase followed by a fairly instantaneous drop in the action max graph (Fig.~\ref{fig::case_2_action}). Usually this will indicate one of two events; first, an annihilation of an instanton/anti-instanton pair; or second, an indication that an object has \emph{fallen through the lattice}.  Features in the topological charge density graph (Fig.~\ref{fig::case_2_topological_charge_density}) suggest that rather than an annihilation, as would be signified by a distinct drop, something else is happening.

Further investigation takes the form of computing the topological charge density at 3 consecutive levels of cooling.  The field data is loaded into the visual tool so that it can be investigated in parallel.  Again, physicists instinctively open the space-like configuration ($xyz$) to search for highlighted minima and maxima.  Upon opening the relevant time slice for all three cooling slices, the most immediate feature is a distinct change in the contour tree for each volume.  The slices before the peak show a relatively large maxima beyond the noise at the centre of the data set; however, the data on the downward slope instead shows that the maxima has disappeared to be replaced by a minima of roughly equal magnitude.

Visual inspection allows the physicists to look at the phenomena from different angle and formulate hypothesis about its provenance and nature.  As can be seen in Fig.~\ref{fig::case2_main}, this irregularity is characterised by a continual deformation of the object relating to a peak in topological charge density.  Additionally, the minima in the field appears to exist in the same position on the lattice, whereas there are no objects of the relevant isovalue in the earlier slices.  An exact reason for this is not yet known, but physicists that have witnessed it were able to reflect upon and develop several possible hypothesis.

%}
%{}

\end{document}